\documentclass[review]{elsarticle-post}

\usepackage{lineno,hyperref}
\modulolinenumbers[5]

\usepackage{amsmath}
\usepackage{amsfonts}
\usepackage{amsbsy}

\usepackage{color}
\usepackage{ulem}

\journal{Journal of Computational Physics}

\newcommand\pd{\partial}
\newcommand\Rey{\mbox{\textit{Re}}}  
\renewcommand{\vec}[1]{\boldsymbol{\mathrm{#1}}} 
\newcommand{\ten}[1]{\overline{\overline{\boldsymbol{\mathrm{{#1}}}}}}
\providecommand\bnabla{\boldsymbol{\nabla}}
\newcommand{\scr}[1]{\mathrm{#1}}
\newcommand{\order}[1]{\mathcal{O} \left( #1\right)}

\newcommand{\red}[1]{{\color{red}#1}}

\graphicspath{{img/}}

\begin{document}

\begin{frontmatter}

\title{A stable fluid-structure-interaction solver for low-density rigid bodies using the immersed boundary projection method}

\author[KTHaddress]{U\v{g}is~L\={a}cis\corref{mycorrespondingauthor}}
\ead{ugis.lacis@mech.kth.se}

\author[FLRaddress]{Kunihiko~Taira}
\ead{ktaira@fsu.edu}

\author[KTHaddress]{Shervin~Bagheri}
\ead{shervin.bagheri@mech.kth.se}

\cortext[mycorrespondingauthor]{Corresponding author. Tel.: +46 8 790 71 77}
\address[KTHaddress]{Linn\'{e} Flow Centre, KTH Mechanics, 10044 Stockholm, Sweden}
\address[FLRaddress]{Department of Mechanical Engineering and Florida Center for Advanced Aero-Propulsion, Florida State University, Tallahassee, Florida 32310, USA}

\begin{abstract}
Dispersion of low-density rigid particles with complex geometries is ubiquitous in both natural and industrial environments. We show that while explicit methods for coupling the incompressible Navier-Stokes equations and Newton's equations of motion are often sufficient to solve for the motion of cylindrical particles with low density ratios, for more complex particles -- such as a body with a protrusion --  they become unstable. We present an implicit formulation of the coupling between rigid body dynamics and fluid dynamics within the framework of the immersed boundary projection method. Similarly to previous work on this method, the resulting matrix equation in the present approach is solved using a block-LU decomposition. Each step of the block-LU decomposition is modified to incorporate the rigid body dynamics. We show that our method achieves second-order accuracy in space and first-order in time (third-order for practical settings), only with a small additional computational cost to the original method. Our implicit coupling yields stable solution for density ratios as low as $10^{-4}$. We also consider the influence of fictitious fluid located inside the rigid bodies on the accuracy and stability of our method.
\end{abstract}


\end{frontmatter}

\linenumbers

\section{Introduction}

During recent years, the original immersed boundary (IB) method~\cite{peskin1972flow} has been extensively developed and gained popularity due to the ability to handle the interaction of objects of complex geometries with fluids. The key feature of the IB method is that the underlying Eulerian grid does not need to be body conforming. The application of the method ranges from fundamental problems of solid particle suspensions~\cite{Uhlmann_JCP_2005,kempe2012improved,Breugem_JCP_2012}, to natural and industrial problems of complex and elastic geometries~\cite{peskin1977numerical,ye1999accurate,kim2001immersed,tseng2003ghost,griffith2012immersed,borazjani2013fluid}.

A number of studies~\cite{Causin20054506,Forster20071278,Borazjani20087587} have reported difficulties with numerical convergence in fluid-structure interaction (FSI) problems, when 
the fluid force acting on the solid dominates over the solid inertial force,
for example, in blood flow through flexible arteries, as discussed by Baek and Karniadakis~\cite{Baek2012629}.
Often the numerical instabilities are attributed to the added mass component. In the IB framework convergence problems have been reported by Borazjani et al.~\cite{Borazjani20087587}. 

Numerical instabilities 
are also present in particulate flows.
Uhlman~\cite{Uhlmann_JCP_2005} developed an efficient direct forcing IB method that describes the coupling between the Newton's equations of motion and the incompressible Navier-Stokes equations for many particles. However, due to the assumption that the fluid inside the particle moves as a solid body, the method suffers from stability issues for density ratio between solid and fluid
below $\rho = \rho_s/\rho_f < 1.2$ (for spheres),
where $ \rho_s$ is the density of the solid and $ \rho_f$ is the density of the fluid. The method was later improved by Kempe and Fr\"{o}hlich~\cite{kempe2012improved} by taking the motion of the fluid inside the particle explicitly into account; they were able to reduce the critical density ratio to $\rho \approx 0.3$. Same approach was also used by Breugem~\cite{Breugem_JCP_2012}.
All of these approaches use explicit coupling (so called weak coupling) between the rigid body and the fluid. The works by Uhlman~\cite{Uhlmann_JCP_2005}, Kempe and Fr\"{o}hlich~\cite{kempe2012improved} and Breugem~\cite{Breugem_JCP_2012} do not consider non-spherical bodies for which the
fluid forces acting on the solid
can be significantly larger.
For such bodies, their algorithms are likely to be stable only for heavier particles with higher $\rho$.

Some studies have investigated the dynamics of more complex (non-spherical) particle geometries~\cite{Borazjani20087587,zheng2010coupled,yang2015non,Gibou20123246}.
Zheng et al.~\cite{zheng2010coupled} investigated human phonation numerically. To solve the motion of human vocal cords and interaction with surrounding air, they have developed a numerical method, in which a sharp-interface IB method is explicitly coupled with-finite difference Navier-Stokes solver. They show that the limiting density ratio, for which the method becomes unstable, is $\rho > 0.25$, which is similar as in the work by Kempe and Fr\"{o}hlich~\cite{kempe2012improved}. They also have derived the necessary time step in order to resolve the motion of vocal cords and preserve stability.
Borazjani et al.~\cite{Borazjani20087587} compared weak coupling (WC) and strong coupling (SC) algorithms within the IB framework for different problems. The SC in their method is ensured using Gauss-Seidel-like iterations within each time step. They noticed that the WC algorithm becomes unstable, when
the mass of the solid structure is reduced below some critical value.
For certain problems they noticed that the SC algorithm also suffers from a similar drawback.
A relaxation scheme for the inner iterations was implemented to overcome the problem, but with increased number of iterations and added computational cost.

Yang and Stern~\cite{yang2015non} have very recently presented a strongly-coupled non-iterative method using fractional step approach to solve Navier-Stokes equations coupled with sharp-interface direct-forcing IB method. They implemented a SC algorithm by introducing an intermediate step in a 
non-inertial reference frame, following the motion of solid body. Yang and Stern demonstrate improved stability properties with stable simulations down to density ratio $\rho \approx 0.1$, which is a significant improvement over work by Kempe and Fr\"{o}hlich~\cite{kempe2012improved}. The derived formulation does not require any iterations within each time step and thus reduces the computational cost.
Another non-iterative method is proposed by Gibou and Min~\cite{Gibou20123246},
which is similar to algorithm described in the current paper.
They advance both fluid and solid through intermediate states, and impose the interaction between fluid and solid during the projection step.
In their work, the solution steps of the projection method for rigid body are deduced empirically by mimicking the fluid solution steps, while we use a more rigorous approach to derive solution steps for rigid body dynamics by a block-LU decomposition.

An alternative
way to achieve better stability properties in  
simulations at low particle densities is including some information of added mass in the computational method. For example, Eldredge~\cite{eldredge2008dynamically} has shown that the FSI computation can be stabilized if the added mass matrix is computed explicitly and added to the body inertia. Furthermore, Wang and Eldredge~\cite{wang2015strongly} have used some information about the added mass to arrive with relaxation factor, which leads to stable simulations.

In the present method, we discretize the system of equations following the approach by Taira and Colonius~\cite{Taira_JCP_2007} and form a discrete linear system of equations. We then decompose the system using a block-LU decomposition, which gives us the prediction step for both fluid and solid body motion, the modified Poisson equation for the dynamic interaction force between fluid and solid, and the projection step for enforcing the interaction of the solid and fluid.
 
Wang and Eldredge~\cite{wang2015strongly} have developed a numerical method in which the null-space fluid solver of Colonius and Taira~\citep{colonius2008fast} is iteratively coupled with general equations for rigid body dynamics. The null-space based IB method~\citep{colonius2008fast} is an extension to the original IB projection method~\cite{Taira_JCP_2007}. Our present FSI solver
employs s direct solver for a positive-definite algebraic system, based on the block-LU decomposition in line with the original fractional step method~\cite{Perot_JCP_1993}, thus eliminating the
need for any iterations within a single time step. This approach is illustrated using a special case of rigid body dynamics (non-deformable objects), while allowing extension to 
deformable, infinitely thin, open filaments and sheets.
In addition, the current method gives direct access to the pressure field, which is useful in many applications.

We characterize the stability properties of our method on a vortex-induced-vibration (VIV) problem for two
particles -- a circular cylinder with and without a splitter plate clamped to the rear end.
The current method is shown to be stable for solving the flow and body dynamics for both bodies for density ratios as low as $10^{-4}$.

In section~\ref{sec:govern-eq}, we discuss the general governing equations for the physical problem of interest. In section~\ref{sec:orig-ibpm}, we describe the basic elements of the IB projection method by Taira and Colonius~\cite{Taira_JCP_2007}. In section~\ref{sec:current-meth}, we present our extension to the IB projection method for FSI problems with rigid bodies. We discuss Newton's equations of motion and couple them with the IB projection method using both explicit (WC) and implicit formulation (SC). We formulate the SC scheme in matrix form and decompose it using a block-LU decomposition. In section~\ref{sec:validation}, we show the convergence properties of the current method.
We also present results of a freely falling and rising circular cylinder and a neutrally buoyant circular cylinder in shear flow and compare our findings with literature. In section~\ref{sec:stab-tests}, we demonstrate the stability properties of our method for the VIV problem mentioned previously.
In section~\ref{sec:fict-fluid}, we investigate the effect of fictitious fluid inside the particle on numerical stability.
Finally,
we draw conclusions in section~\ref{sec:conclusions}.
 In ``Appendix. Stability of the implicit coupling for massless particles'', we present a modification of the present method, which is stable for limiting case of massless particles. In ``Appendix. Designing a parallel Poisson solver by using the block-LU decomposition'', we show a design of a parallel algorithm for solution of Poisson equation, which does not depend on domain decomposition.

\section{Governing equations of a rigid body motion in a fluid} \label{sec:govern-eq}

\begin{figure}[t!]
  \centering
  \includegraphics{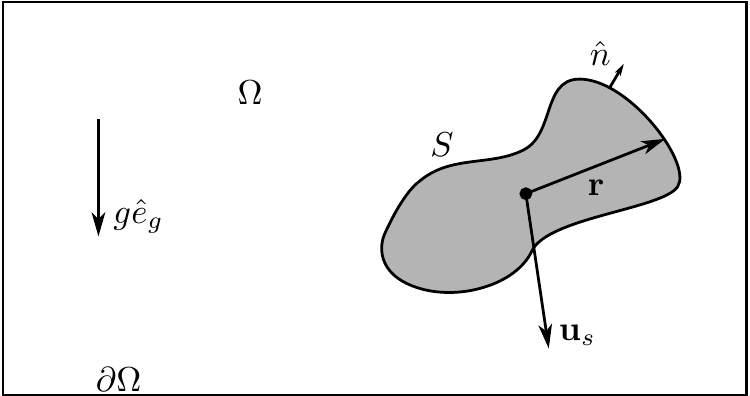} 
  \caption{Schematic representation of fluid domain $\Omega$ and a solid particle.}
\label{fig:domain-setup}
\end{figure}

We consider a general solid body (represented with gray body in Fig.~\ref{fig:domain-setup}) immersed in a viscous, incompressible fluid. We denote the fluid domain with $\Omega$, the outer boundary with $\pd \Omega$ and the boundary at solid body with $S$. The whole system can be subject to gravitational acceleration\footnote{The formulation holds also for uniform background acceleration.} $\vec{g}$. The solid body moves with velocity $\vec{u}_s$ under the influence of gravity and contact forces from the fluid.

This configuration is governed by a system of non-dimensional Navier-Stokes equations and Newton's equations of motion
\begin{align}
\frac{\pd \vec{u}}{\pd t} + \left( \vec{u} \cdot \bnabla \right) \vec{u} & = - \bnabla \tilde{p} + \frac{1}{\Rey} \nabla^2 \vec{u} + g \hat{e}_g & \mbox{in } \Omega, \label{eq:NS1-orig} \\
\bnabla \cdot \vec{u} & = 0  & \mbox{in } \Omega, \label{eq:NS2-orig} \\
\vec{u} & = \vec{u}_s + \vec{\omega}_s \times \vec{r} & \mbox{on } S, \label{eq:orig-bc1} \\
a \vec{u} + b \frac{\pd \vec{u}}{\pd \hat{n}} & = c & \mbox{on } \pd \Omega, \label{eq:orig-bc2} \\
\frac{d\vec{u}_s}{d t} & = \frac{1}{\rho V_s} \oint_{S} \ten{\tau} \cdot \hat{n}\,\mathit{dS} + \left( 1 - \frac{1}{\rho} \right) g \hat{e}_g, & \label{eq:orig-ND1} \\
\frac{d \left( \ten{I}_s \cdot \vec{\omega}_s \right)}{d t} & = \frac{1}{\rho} \oint_{S} \vec{r} \times \left(\ten{\tau} \cdot \hat{n} \right) \mathit{dS}, & \label{eq:orig-ND2}
\end{align}
where equations (\ref{eq:orig-bc1}--\ref{eq:orig-bc2}) are the boundary conditions on the physical and computational domains, and $a$, $b$ and $c$ are known parameters.
The vector field $\vec{u}$ is the fluid velocity field, the scalar field $\tilde{p}$ is the fluid pressure, $Re$ (defined later) is the Reynolds number, and $\ten{\tau}$ is the fluid stress tensor. In Newton's equations of motion, $\vec{u}_s$ is the translation velocity located at the center of mass for the given body, $\vec{\omega}_s$ is the angular velocity of the center of mass, $\vec{r}$ is the radius from the center of mass to the surface of the body or some other point in fluid, $V_s = \int dV$ is dimensionless volume, and $\ten{I}_s$ is dimensionless second-rank moment of inertia tensor with components $I_s^{ij} = \int \left( r_k r_k \delta_{ij} + r_i r_j \right) \mathit{dV}$.

\section{Immersed boundary projection method} \label{sec:orig-ibpm}

We start with writing the incompressible Navier-Stokes equations (\ref{eq:NS1-orig}-\ref{eq:NS2-orig})
with boundary conditions (\ref{eq:orig-bc1}--\ref{eq:orig-bc2}) in IB formulation. The solid body is replaced with fluid and volume forcing, to mimic the boundary conditions from the solid body. The system of equations is given by
\begin{align}
\frac{\pd \vec{u}}{\pd t} + \left( \vec{u} \cdot \bnabla \right) \vec{u} & = - \bnabla p + \frac{1}{\Rey} \nabla^2 \vec{u} + \int_S \vec{F\left(\mathcal{L}\right)} \delta\left( \vec{\mathcal{L}} - \vec{x} \right)\,\mathit{dS} & \mbox{in } \Omega, \label{eq:NS1-ibm} \\
\bnabla \cdot \vec{u} & = 0  & \mbox{in } \Omega, \label{eq:NS2-ibm} \\
\vec{u\left(\mathcal{L}\right)} & = \int_{\Omega} \vec{u\left(\vec{x}\right)} \delta\left( \vec{x} - \vec{\mathcal{L}} \right)\,\mathit{dV} = \vec{u}_s + \vec{\omega}_s \times \vec{\tilde{\mathcal{L}}}, \label{eq:NS3-ibm} \\
\vec{u} & = 0 & \mbox{on } \pd \Omega, \label{eq:NS4-ibm}
\end{align}
where $\vec{F}$ is the surface Lagrangian force density of the IB method (applied only on solid body surface or Lagrangian points), $\vec{\mathcal{L}}$ is a vector directed to the IB Lagrangian points, $\vec{\tilde{\mathcal{L}}}$ is a vector directed to the IB Lagrangian points from the center of solid body, $\vec{x}$ is the coordinate vector in Eulerian space, and $\delta$ is the Dirac delta function. The effect of gravity in the Navier-Stokes equations from now on is incorporated in the modified pressure field $p$. Without the loss of generality, we apply no-slip boundary condition ($a \neq 0$, $b = c = 0$) on $\pd \Omega$. The fluid solver used in the current work is the finite volume IB projection method on a non-uniform staggered grid~\cite{ferziger2002computational}. The delta function is approximated using $3$-cell discrete delta function developed by Roma et al.~\cite{roma1999adaptive}.
The diffusive term is integrated with the second-order implicit Crank--Nicolson method; the non-linear advective term is advanced in time with the second-order explicit Adams--Bashforth method. Consequently, equations (\ref{eq:NS1-ibm}--\ref{eq:NS4-ibm}) become
\begin{align}
\frac{u^{n+1}-u^{n}}{\Delta t} + \left[ \frac{3}{2} \hat{N} \left(u^{n}\right) - \frac{1}{2}\hat{N}\left(u^{n-1}\right) \right] & = - \hat{G} \phi^{n+1/2} + \frac{1}{2 \Rey} \hat{L} \left( u^{n+1} + u^{n} \right) +  \nonumber \\
 & + \hat{H}_n f^{n+1/2}  + \hat{bc}_1,  \label{eq:num-NSeq-1} \\
\hat{D} u^{n+1} & = 0 + \hat{bc}_2, \\
\hat{E}_n u^{n+1} & = 	u^{n}_s + \omega^{n}_s \times \tilde{\mathcal{L}}^{n}, \label{eq:num-NSeq-3}
\end{align}
where $\hat{N}\left(\vec{u}\right)$ is the non-linear (or advection) operator, $\hat{G}$ is the gradient operator, $\hat{L}$ is the Laplace operator, $\hat{D}$ is the divergence operator, $\hat{H}_n$ is the spreading (regularization) operator, $\hat{E}_n$ is the interpolation operator, $\phi^{n+1/2}$ is the discrete pressure, $f^{n+1/2}$ is the discrete IB forcing vector, and $\hat{bc}_1$ and $\hat{bc}_2$ are boundary conditions associated with the momentum and continuity equations, respectively. Note that we have introduced the subscript $n$ to denote the time level used by the interpolation and spreading operators. This illustrates that in the case of a moving body we do not know the position of Lagrangian points \textit{a priori}, therefore one option is to use positions at previous time step $n$. In such way, we obtain a time-lagged interpolation, where the interpolation operator at time level $n$ acts on the velocity flux at time level $n+1$. We will discuss this issue in further details in section~\ref{sec:current-meth}.

The divergence and gradient matrices can be made to consist of only $1$ and $-1$ by introducing diagonal scaling matrices $\hat{M}$ and $R$, i.e. $G = \hat{M}\hat{G}$ and $D = \hat{D} R^{-1}$ ($G = -D^T$), as detailed in Taira and Colonius~\cite{Taira_JCP_2007}.
Using a similar transformation, the interpolation matrix $\hat{E}$ and spreading matrix $\hat{H}$ can be made transpose of each other. 
We introduce an implicit operator $A = \hat{M} \left[I/\Delta t - \hat{L} / ( 2 Re ) \right] R^{-1}$, which contains the terms in front of the unknown velocity field at the time level $n+1$. Adopting the scaling above, equations (\ref{eq:num-NSeq-1}--\ref{eq:num-NSeq-3}) of the IB formulation can be written in algebraic form as
\begin{equation}
\left( \begin{array}{ccc}
A & G & E_n^T \\
G^T & 0 & 0 \\
E_n & 0 & 0	
\end{array} \right) \left( \begin{array}{c}
q^{n+1} \\
\phi^{n+1/2} \\
\tilde{f}^{n+1/2}
\end{array} \right) = \left( \begin{array}{c}
r^n \\
0 \\
u_s^{n}
\end{array} \right) + \left( \begin{array}{c}
bc_1 \\
- bc_2 \\
0
\end{array} \right), \label{eq:origTaira-full-mat}
\end{equation}
where $q^{n+1} = R u^{n+1}$ is the velocity flux vector, $\tilde{f}^{n+1/2}$ is a rescaled IB forcing vector and $u_s^{n}$ is prescribed velocity of IB Lagrangian points. The IB forcing $f^{n+1/2}$ has been rescaled to obtain symmetry between blocks $(1,3)$ and $(3,1)$ as shown by Taira and Colonius~\cite{Taira_JCP_2007}.
The resulting matrix can be decomposed in the same way as performed by Perot~\cite{Perot_JCP_1993} using a $N$th-order temporal approximation of the inverse of $A$, i.e.
\begin{equation}
A^{-1} \approx B^N = \sum_{i=1}^N \frac{\Delta t^i}{2^{i-1}} \left( M^{-1} L \right)^{i-1} M^{-1}, \label{eq:proj-meth-BN-expansion}
\end{equation}
where $M = \hat{M} R^{-1}$ and $L = \hat{M} \hat{L} R^{-1}$. The approximation of the inverse $B^{N}$ has $\mathcal{O}\left( \Delta t^N \right)$ truncation error. Next, we employ the block-LU decomposition and arrive with three steps
to solve the problem, i.e.
\begin{align}
A q^* & = r^n + bc_1, \label{eq:taira-proj1} \\
\left( \begin{array}{cc}
G^T B^N G & G^T B^N E_n^T \\
E_n B^N G & E_n B^N E_n^T
\end{array} \right) \left( \begin{array}{c}
\phi^{n+1} \\ \tilde{f}^{n+1/2}
\end{array} \right) & = \left( \begin{array}{c}
G^T \\ E_n
\end{array} \right) q^* - \left( \begin{array}{c}
-bc_2 \\ u_s^{n}
\end{array} \right), \label{eq:taira-proj2} \\
q^{n+1} = q^* & - B^N \left( G \phi^{n+1} + E_n^T \tilde{f}^{n+1/2} \right), \label{eq:taira-proj3}
\end{align}
where equation (\ref{eq:taira-proj1}) is the so-called prediction step ($q^*$ is the intermediate velocity flux), equation (\ref{eq:taira-proj2}) is the pressure-Poisson step, and equation (\ref{eq:taira-proj3}) is the projection step.
The matrix $A$ in equation (\ref{eq:taira-proj1}) and the block matrix on the left-hand-side of equation (\ref{eq:taira-proj2}) are positive definite;
hence the conjugate gradient method or the Cholesky factorization can be used to solve the linear systems
in an efficient manner.
Note that a modified finite volume scheme near the boundary may make the Laplacian $L$ non-symmetric. One may symmetrize $L$ by a similarity transform.

\section{Numerical treatment of rigid body motion in fluid} \label{sec:current-meth}

\subsection{Newton's equations of motion}

When the rigid body dynamics are not known \textit{a priori}, it must be solved using Newton's equations of motion. These equations in dimensionless form are repeated for convenience
\begin{align}
\frac{d\vec{u}_s}{d t} & = \frac{1}{\rho V_s} \oint_{S} \ten{\tau} \cdot \hat{n}\,\mathit{dS} + \left( 1 - \frac{1}{\rho} \right) g \hat{e}_g, & \label{eq:IB-Newt-rep1} \\
\frac{d\left( \ten{I}_s \cdot \vec{\omega}_s \right)}{d t} & = \frac{1}{\rho} \oint_{S} \vec{r} \times \left(\ten{\tau} \cdot \hat{n} \right)\,\mathit{dS}. & \label{eq:IB-Newt-rep2}
\end{align}
The Newton's equations of motion in the IB framework pose a difficulty, because there is a fluid inside the solid body.
One may consider the solid body as 
a region with fictitious boundary $S^+$, which encompasses the IB forcing and the fluid within. Stresses over that given surface can be related to the flow field inside the volume and the volume forcing by
\begin{align}
 \oint_{S^{+}} \ten{\tau} \cdot \hat{n}\,\mathit{dS} & = - \int_{V^+} \tilde{\vec{F}}\,\mathit{dV} + \frac{d}{dt} \int_{V^+} \vec{u}\,\mathit{dV}, \label{eq:fluid-stress-int1} \\
 \oint_{S^{+}} \vec{r} \times \left(\ten{\tau} \cdot \hat{n} \right)\,\mathit{dS} & = - \int_{V^+} \vec{\tilde{\mathcal{L}}} \times \tilde{\vec{F}}\,\mathit{dV} + \frac{d}{dt} \int_{V^+} \vec{r} \times \vec{u}\,\mathit{dV}, \label{eq:fluid-stress-int2}
\end{align}
where  $\tilde{\vec{F}} = \int_S \vec{F\left(\mathcal{L}\right)} \delta\left( \vec{\mathcal{L}} - \vec{x} \right) \mathit{dS}$ is the IB volume force density (applied on Eulerian grid), and the integration is over the fluid volume $V^+$
encompassed by surface $S^{+}$.
We note that the force density $\tilde{\vec{F}}$ is non-zero only in a thin region close to the surface of the body.
We assume that the spatial support of the discrete delta function is compact and consider the spatially limiting case as $S^+ \rightarrow S$ and $V^+ \rightarrow V$, where $S$ and $V$ is the surface and volume of actual solid body, respectively.
Following the suggestion by Kempe and Fr\"{o}hlich~\cite{kempe2012improved}, we evaluate the linear and angular acceleration of the fluid inside the particle as separate terms. A simpler approach suggested by Uhlman~\cite{Uhlmann_JCP_2005} is analyzed in section~\ref{sec:fict-fluid}. Inserting equations (\ref{eq:fluid-stress-int1}--\ref{eq:fluid-stress-int2}) in (\ref{eq:IB-Newt-rep1}--\ref{eq:IB-Newt-rep2}), we arrive with Newton's equations of motion in the IB framework
\begin{align}
\rho V_s \frac{d\vec{u}_s}{d t} & = - \oint_{S} \vec{F}\,\mathit{dS} + \frac{d}{dt} \int_V \vec{u}\,\mathit{dV} + V_s \left( \rho - 1 \right) g \hat{e}_g, \label{eq:Newt1-ibm} & \\
\rho \frac{d \left( \ten{I}_s \cdot \vec{\omega}_s \right)}{d t} & = - \oint_{S} \vec{\tilde{\mathcal{L}}} \times \vec{F}\,\mathit{dS} + \frac{d}{dt} \int_V \vec{r} \times \vec{u}\,\mathit{dV}, \label{eq:Newt2-ibm} &
\end{align}
where we have substituted the
volume force on the Eulerian grid with surface force on the Lagrangian grid by integrating over the delta function.
Note that the first term on the right-hand side of equation (\ref{eq:Newt1-ibm}) and (\ref{eq:Newt2-ibm}) is integral over the surface of the solid body, which is later discretized using Lagrangian points, whereas the second term is volume integral over the fluid volume, which is later discretized using Eulerian mesh. We believe this to be the best representation of the equations in our method, since the unknown forcing is defined on the Lagrangian points, while unknown flow field is defined on the Eulerian mesh, as described is section \ref{sec:orig-ibpm}.
	
While we only explain the detailed structure of matrices in two dimensions, the extension to three dimensions is straightforward. The moment of inertia tensor in a two-dimensional setting simplifies to a scalar, time independent constant $I_s = \int \vec{r}^2\,\mathit{dV}$. We introduce the solid body velocity variables $u_s$, $v_s$ (translation in $x$ and $y$ directions, respectively) and $\omega_s$ (angular velocity) of center of mass for the rigid body. We can then discretize the Newton's equations of motion (\ref{eq:Newt1-ibm}--\ref{eq:Newt2-ibm}) as
\begin{align}
\rho V_s \frac{u_s^{n+1} - u_s^{n}}{\Delta t}   = & - \sum_{i \in \mathcal{L}} \tilde{f}_{x_i} + dQ_{x}^{n+1/2}  + V_s \left( \rho - 1 \right) g_{x}, \label{eq:Newt-ibm-disc1} \\
\rho V_s \frac{v_s^{n+1} - v_s^{n}}{\Delta t}   = & - \sum_{i \in \mathcal{L}} \tilde{f}_{y_i} + dQ_{y}^{n+1/2}  + V_s \left( \rho - 1 \right) g_{y}, \label{eq:Newt-ibm-disc2} \\
\rho I_s \frac{\omega_s^{n+1} - \omega_s^{n}}{\Delta t}   = & - \sum_{i \in \mathcal{L}} \left( \tilde{\mathcal{L}}^{n}_{x_i} \tilde{f}_{y_i} - \tilde{\mathcal{L}}^{n}_{y_i} \tilde{f}_{x_i} \right) + dQ_{\omega}^{n+1/2}, \label{eq:Newt-ibm-disc3}
\end{align}
where $\tilde{\mathcal{L}}^{n}_{x_i} = \mathcal{L}^{n}_{x_i} - x^{n}_c$ and $\tilde{\mathcal{L}}^{n}_{y_i} = \mathcal{L}^{n}_{y_i} - y^{n}_c$ are the relative coordinates of the body surface point $i$ measured from position of the center of mass $(x^{n}_c,  y^{n}_c)$, and $\tilde{f}_{x_i}$ and $ \tilde{f}_{y_i}$ are the IB forcing components in $x$ and $y$ directions on the surface point $i$, where $i \in \left(1, nb\right)$. Here, $nb$ is number of Lagrangian surface points. 
After the solution for the body velocity is found, the coordinates of Lagrangian points are updated as
\begin{align}
\mathcal{L}^{n+1}_{x_i}  = & \mathcal{L}^{n}_{x_i} + \Delta t \left( u_s^{n+1} - \omega^{n+1}_s \tilde{\mathcal{L}}^{n}_{y_i} \right), \label{eq:NewtCoord-ibm-disc1} \\
\mathcal{L}^{n+1}_{y_i}  = & \mathcal{L}^{n}_{y_i} + \Delta t \left( v_s^{n+1} + \omega^{n+1}_s \tilde{\mathcal{L}}^{n}_{x_i} \right) , \label{eq:NewtCoord-ibm-disc2}
\end{align}
for all $i$ values from $1$ to $nb$.
Note that we have omitted the time level index for the forcing, since it can have different indices depending on
the particular integration scheme for coupling the equations of motion with the fluid equations (see next section). The terms $g_x$ and $g_y$ are the gravitational acceleration in $x$ and $y$ directions, respectively. We have used
a first-order linear approximation of time derivative for volume integral of fluid inside particle. This derivative is denoted by $dQ_{x}^{n+1/2}$, $dQ_{y}^{n+1/2}$, $dQ_{\omega}^{n+1/2}$ for integrals $\frac{d}{dt} \int_V u\,\mathit{dV}$, $\frac{d}{dt} \int_V v\,\mathit{dV}$ and $\frac{d}{dt} \int_V \left(r_x v - r_y u \right)\mathit{dV}$, where $u$ and $v$ are the flow velocity components in $x$ and $y$ directions, respectively. The time level of the derivative is $n+1/2$; and the derivative depends on two previous flow fields
$u^{n}$ and $u^{n-1}$.
The coefficients in front of integral value at each time level for first and higher-order approximations can be derived as explained by Fornberg~\cite{fornberg1988generation}.
For example, the time derivative approximation $dQ_{x}^{n+1/2}$
can be expanded as
\begin{align}
& dQ_{x}^{n+1/2} ( u^{n}, u^{n-1}) = \frac{Q_{x} u^{n} - Q_{x} u^{n-1}}{\Delta t}, \label{eq:Newt-fluid-int-approx}
\end{align}
where $Q_{x}$ is the second-order midpoint rule
for $u$ velocity component. In some fluid cells, the solid particle fills only a fraction of the cell. Therefore the quadrature in two dimensions takes the form
\begin{equation}
Q_{x} u^{n} = \sum_{ij} u^n_{ij} \alpha_{ij}\, \Delta x\, \Delta y, \label{eq:fluid-int-quad}
\end{equation}
where $\alpha_{ij}$ is the solid volume fraction in the fluid cell $(i,j)$ centered around grid point for fluid velocity component $u^n_{ij}$. Apart from solid volume fraction $\alpha_{ij}$, equation (\ref{eq:fluid-int-quad}) is the standard second-order midpoint rule~\cite{ferziger2002computational}.
As employed by Kempe and Fr\"{o}hlich~\cite{kempe2012improved}, the solid volume fraction $\alpha_{ij}$ can be found by the Heaviside step function $H$ and signed distance function $\zeta$ between a grid point and the surface of the body as
\begin{equation}
\alpha_{ij} = \frac{\sum_{n=1}^{4} - \zeta_n H\left( - \zeta_n \right)}{\sum_{n=1}^{4} \left| \zeta_n \right| },
\end{equation}
where summation is performed over all corners of the fluid cell $(i,j)$. The term $\zeta$ is assumed to be positive outside the body, and negative inside the body. We note that the Heaviside step function is used only as indicator function and is not discretized, contrary to delta function of IB formulation, which is discretized and smoothed.

In order to write the Newton's equations of motion in a more compact form, we introduce a solid body velocity vector,
\begin{equation}
u_B^{n} = \left( \begin{array}{ccc} u_s^{n} & v_s^{n} & \omega_s^{n} \end{array} \right)^T,
\end{equation}
a diagonal matrix of inertia\footnote{This is similar to generalized inertia matrix $\mathbb{H}$ in the work of Wang and Eldredge~\cite{wang2015strongly}.},
\begin{equation}
I_B = \frac{\rho}{\Delta t} \left( \begin{array}{ccc} V_s & 0 & 0 \\
0 & V_s  & 0 \\
0 & 0 & I_s \end{array} \right), \label{eq:diag-mat-inert}
\end{equation}
an inner-fluid integral vector,
\begin{equation}
dQ_B^{n+1/2} = \left( \begin{array}{ccc} dQ_{x}^{n+1/2} & dQ_{y}^{n+1/2} & dQ_{\omega}^{n+1/2} \end{array} \right)^T, \label{eq:fict-fluid-int-all}
\end{equation}
and a gravitational acceleration vector,
\begin{equation}
G_B = \left( \begin{array}{ccc} g_x & g_y & 0 \end{array} \right)^T.
\end{equation}
We use the following arrangement of the IB forcing values,
\begin{equation}
\tilde{f} = \left( \begin{array}{cccccccc} \tilde{f}_{x_{1}} & \ldots & \tilde{f}_{x_{nb-1}} & \tilde{f}_{x_{nb}} & \tilde{f}_{y_{1}} & \ldots & \tilde{f}_{y_{nb-1}} & \tilde{f}_{y_{nb}}  \end{array} \right)^T. \label{eq:IB-force-vec-expl}
\end{equation}
The force summation matrix\footnote{This is equivalent to transpose of rigid-body distribution operator $\mathcal{B}^T$~\cite{wang2015strongly}.} at time level $n$ is defined as
\begin{equation}
N^n_B = - \left( \begin{array}{cccccc}
 1 & \ldots &  1 &  0 & \ldots &  0 \\
 0 & \ldots &  0 &  1 & \ldots &  1 \\
 -\tilde{\mathcal{L}}^n_{y_1} & \ldots &-\tilde{\mathcal{L}}^n_{y_{nb}}
& \tilde{\mathcal{L}}^n_{x_1} & \ldots  & \tilde{\mathcal{L}}^n_{x_{nb}}
\end{array} \right). \label{eq:IB-int-mat}
\end{equation}
Using the above matrices and vectors, we can express the discrete Newton's equations (\ref{eq:Newt-ibm-disc1}--\ref{eq:Newt-ibm-disc3}) in the following algebraic form
\begin{equation}
I_B \left( u_B^{n+1} - u_B^{n} \right) = N^{n+1/2}_B \tilde{f} + dQ_B^{n+1/2} + G_B. \label{eq:Newton-disc-mat}
\end{equation}
Below we describe the explicit and implicit approaches for solving this equation in the framework of the IB projection method.

\subsection{Explicit coupling between fluid and rigid-body dynamics} \label{sec:expl-coup-sch}

We start with describing forward Euler's and Heun's (predictor-corrector) methods for coupling the fluid-structure interaction, in order to compare with fully implicit scheme described in section~\ref{sec:imp-coupl-desc}. 
We denote the fluid solver as a function $\tilde{f}^{n+1/2} = \mathrm{NS} ( u^n_B, \mathcal{L}^n )$, which takes the solid body velocity vector $u^n_B$ and the coordinate vector $\mathcal{L}^n$ as function arguments at time level $n$, and returns the IB forcing $\tilde{f}^{n+1/2}$. Note that we are reintroducing a specific time level $n+1/2$ for the IB forcing $\tilde{f}$ to illustrate the time level used in the coupling. Similarly, we denote the discrete Newton's equations of motion (\ref{eq:Newton-disc-mat}) as a function $u_B^{n+1} = \mathrm{ND} ( \tilde{f}^n  )$, and coordinate update (\ref{eq:NewtCoord-ibm-disc1} -- \ref{eq:NewtCoord-ibm-disc2})
procedure as $\mathcal{L}^{n+1}
 = \mathrm{NC} ( u_B^n  )$.
Then, the Euler forward integration in time for the coupled system can be schematically written as
\begin{align}
\tilde{f}^{n+1/2} & = \mathrm{NS}\left(u^n_B, \mathcal{L}^n\right) \label{eq:rbd-Euler1} \\ 
u_B^{n+1} & = \mathrm{ND}\left( \tilde{f}^n \right), \label{eq:rbd-Euler2} \\
\mathcal{L}^{n+1} & = \mathrm{NC}\left( u_B^n \right), \label{eq:rbd-Euler3}
\end{align}
where $\tilde{f}^n = (\tilde{f}^{n+1/2} + \tilde{f}^{n-1/2})/2$ is an interpolated IB forcing at time level $n$. The outlined time stepping follows closely the definition of the forward Euler's method, i.e.
on the right hand side we only have parameters at time level $n$.

Next, we describe Heun's method (a predictor-corrector method).
For the fluid-structure interaction problem, the first step of the Heun's method is the forward Euler's method discussed before
\begin{align}
\tilde{f}^{*} & = \mathrm{NS}\left(u^n_B, \mathcal{L}^n\right), \label{eq:rbd-Heun1} \\ 
u_B^{*} & = \mathrm{ND}\left( \tilde{f}^{n*} \right), \\
\mathcal{L}^{*} & = \mathrm{NC}\left( u_B^n \right), 
\end{align}
where $u_B^{*}$ and $\mathcal{L}^{*}$ are predictions for the solid body velocity and the coordinates respectively, and $\tilde{f}^{n*} = ( \tilde{f}^{n-1/2} + \tilde{f}^{*})/2$ is the IB forcing at time level $n$ using the predicted value $\tilde{f}^{*}$ for interpolation. Next, the correction step is carried out for the Navier-Stokes equations and the results are used to advance the solid body dynamics
\begin{align}
\tilde{f}^{n+1/2} & = \mathrm{NS}\left(\frac{u^n_B + u^*_B}{2}, \frac{\mathcal{L}^n+\mathcal{L}^*}{2} \right), \\ 
u_B^{n+1} & = \mathrm{ND}\left( \tilde{f}^{n} \right), \\
\mathcal{L}^{n+1} & = \mathrm{NC}\left( u_B^n \right), \label{eq:rbd-Heun6} 
\end{align}
where the IB forcing $\tilde{f}^{n}$ at time level $n$ is interpolated using
 $\tilde{f}^{n+1/2}$ and  $\tilde{f}^{n-1/2}$.
Note that other combinations of time coupling are possible. For example, one could use the intermediate result from Navier-Stokes equations in Newton's equations of motion or vice versa. We adopt the above approach, which resembles the Heun's method as close as possible.

\subsection{Implicit coupling between fluid and rigid body dynamics} \label{sec:imp-coupl-desc}

We formulate implicit coupling (SC) as
\begin{align}
\tilde{f}^{n+1/2} & = \mathrm{NS}\left(u^{n+1}_B, \mathcal{L}^n\right) \label{eq:def-impl1} \\ 
u_B^{n+1} & = \mathrm{ND}\left( \tilde{f}^{n+1/2} \right), \label{eq:def-impl2} \\
\mathcal{L}^{n+1} & = \mathrm{NC}\left( u_B^{n+1} \right), \label{eq:def-impl3}
\end{align}
where we observe that the output from solver (\ref{eq:def-impl1}) is used in (\ref{eq:def-impl2}) and vice versa. Therefore both the Navier-Stokes equations and the Newton's equations of motion have to be solved simultaneously. Note that our SC scheme includes only implicit coupling for solid body velocity $u^{n+1}_B$. There is no straightforward way to include body coordinates $\mathcal{L}^n$ implicitly, because the interpolation and spreading operators $E_n$ and $E^T_n$ are depending on the coordinates of Lagrangian points $\mathcal{L}^n$, which would make the overall system non-linear. Consequently the overall temporal accuracy of the presented method is first order
due to time-lagged interpolation and spreading operations\footnote{The dominant first-order contribution is $\delta\left( \vec{\mathcal{L}}^{n+1/2} - \vec{x} \right) = \delta\left( \vec{\mathcal{L}}^{n} - \vec{x} \right) + \order{\Delta t}$.}.
In addition, since the system of equations now is solved at the same time, the force vector $\tilde{f}$ does not necessarily have to have a time level -- it can be viewed as Lagrange multiplier. Nevertheless, we choose to keep the time level notation to be consistent with explicit methods.

The coupled Navier-Stokes and Newton's equations can be written in the following algebraic form
\begin{equation}
\left( \begin{array}{cccc}
A & 0 & G & E_n^T \\
0 & I_B & 0 & N^n_B \\
G^T & 0 & 0 & 0 \\
E_n & (N_B^n)^T & 0 & 0	
\end{array} \right) \left( \begin{array}{c}
q^{n+1} \\
u_B^{n+1} \\
\phi^{n+1/2} \\
\tilde{f}^{n+1/2}
\end{array} \right) = \left( \begin{array}{c}
r^n \\
r^{n}_B \\
0 \\
\Delta u_B^{n+1}
\end{array} \right) + \left( \begin{array}{c}
bc_1 \\
0 \\
- bc_2 \\
0
\end{array} \right). \label{eq:full-mat-Newton}
\end{equation}
Here the Newton's equations of motion (\ref{eq:Newton-disc-mat}) are the second block-row and $r^{n}_B$ contains all known terms of the equations of rigid-body motion, i.e. the velocity at previous time step, the external forces at time level $n+1/2$, and the derivative of velocity field inside the body at time level $n+1/2$.
The coupling between rigid-body dynamics and fluid dynamics is ensured using IB forcing $\tilde{f}^{n+1/2}$ both as volume forcing in Navier-Stokes equations (\ref{eq:origTaira-full-mat}) and as surface forcing for total force integral in Newton's equations of motion (\ref{eq:Newton-disc-mat}). Furthermore, the coupling appears through the prescribed velocity at the solid body boundary (\ref{eq:num-NSeq-3}), in which the velocity value at each boundary point is constructed using the unknown solid body velocity $u^{n+1}_B$
\begin{equation}
E_n q^{n+1} =  - (N_B^n)^T u^{n+1}_B + \Delta u_B^{n+1},
\end{equation}
where we have introduced $\Delta u_B^{n+1}$ as a prescribed difference between flow velocity at the boundary and the velocity of the body. For the no slip condition $\Delta u_B^{n+1} = 0$, whereas non-zero values correspond to
a prescribed slip or penetration velocity, which results in
some
force exerted from the particle onto the fluid in order to match the prescribed slip or penetration velocity exactly. If the velocity difference is time dependent, the time level $n+1$ is selected to match the time level of solid body velocity.

Next, we can decompose the system (\ref{eq:full-mat-Newton}) using a block-LU decomposition,
\begin{equation}
\left( \begin{array}{cc}
\tilde{A} & \tilde{Q} \\
\tilde{Q}^T & 0
\end{array} \right) = \left( \begin{array}{cc}
\tilde{A} & 0 \\
\tilde{Q}^T & - \tilde{Q}^T \tilde{B}^N \tilde{Q}
\end{array} \right) \left( \begin{array}{cc}
I & \tilde{B}^N \tilde{Q} \\
0 & I
\end{array} \right), \label{eq:perot-decomp}
\end{equation}
where we have defined
\begin{equation}
\tilde{A} = \left( \begin{array}{cc}
A & 0  \\
0 & I_B
\end{array} \right) \ \ \mbox{}\ \  \tilde{Q} = \left( \begin{array}{cc}
G & E_n^T \\
0 & N^n_B
\end{array} \right). \label{eq:mat-Newton-blocks}
\end{equation}
The matrix $\tilde{A}$ is symmetric and positive definite, since both $A$ and $I_B$ are symmetric and positive definite.
The approximate inverse $\tilde{B}^N$ of order $N$ is
\begin{equation}
\tilde{B}^N = \left( \begin{array}{cc}
B^N & 0  \\
0 & I_B^{-1}
\end{array} \right), \label{eq:matrix-tildeBn}
\end{equation}
where $B^{N}$ is the $N$th-order approximation of $A^{-1}$ from equation (\ref{eq:proj-meth-BN-expansion}).

To summarize the method, let us list the three steps for IB projection method with implicit solver for rigid body dynamics;
\begin{enumerate}[(I)]
   \item prediction step
\begin{equation}
\left( \begin{array}{cc}
A & 0  \\
0 & I_B
\end{array} \right)  \left( \begin{array}{c}
q^{*} \\
u^{*}_B
\end{array} \right) = \left( \begin{array}{c}
r^{n} + bc_1 \\
r^{n}_B
\end{array} \right), \label{eq:NewtMat-proj1}
\end{equation}
   \item modified pressure Poisson solver
\begin{align}
& \left( \begin{array}{cc}
G^T B^N G & G^T B^N E_n^T \\
E_n B^N G & E_n B^N E_n^T + (N_B^n)^T I_B^{-1} N^n_B
\end{array} \right) \left( \begin{array}{c}
\phi^{n+1/2} \\ \tilde{f}^{n+1/2}
\end{array} \right) \nonumber \\
= & \left( \begin{array}{c}
G^T q^* + bc_2 \\ E_n q^* + (N_B^n)^T u_B^{*} - \Delta u_B^{n+1}
\end{array} \right), \label{eq:NewtMat-proj2}
\end{align}
  \item projection step (to enforce incompressibility and rigid body dynamics)
\begin{equation}
\left( \begin{array}{c}
q^{n+1} \\
u_B^{n+1} \end{array} \right) =  \left( \begin{array}{c}
q^* \\
u_B^* \end{array} \right) - \left( \begin{array}{ccc}
 B^N G \phi^{n+1/2} & + & B^N E_n^T \tilde{f}^{n+1/2} \\
  & & I_B^{-1} N^n_B \tilde{f}^{n+1/2} \end{array} \right). \label{eq:NewtMat-proj3}
\end{equation}
\end{enumerate}
It is noteworthy that the interaction between the solid body and the fluid is computed by modifying the Poisson matrix (\ref{eq:NewtMat-proj2}) using block matrix
$(N_B^n)^T I_B^{-1} N^n_B$
and by modifying the right-hand side using the predicted solid body velocity $(N_B^n)^T u_B^{*}$. Therefore, the added computational cost to the original method by Taira and Colonius~\cite{Taira_JCP_2007} is minimal. The size of modified pressure Poisson algebraic system is identical, while the algebraic system of prediction step is complemented only by $3$ or $6$ rows and columns for a single body in a two or three-dimensional setting, respectively (corresponds to a number of degrees of freedom for the rigid body).

It is straightforward to modify the method to include multiple bodies. In the case of $m$ bodies, Lagrangian points for all considered bodies must be assembled, and the corresponding velocity array, force array and diagonal matrix of inertia would be extended as
\begin{equation}
\hat{u}_B = \left( \begin{array}{c}
u_B^1 \\
u_B^2 \\
\vdots \\
u_B^m
\end{array} \right), \ \ \ \ \hat{\tilde{f}} = \left( \begin{array}{c}
\tilde{f}^1 \\
\tilde{f}^2 \\
\vdots \\
\tilde{f}^m
\end{array} \right), \ \ \ \ \hat{I}_B = \left( \begin{array}{cccc}
I_B^1 & 0 & \cdots & 0 \\ 
0 & I_B^2 & \cdots & 0 \\
\vdots & \vdots & \ddots & \vdots \\
0 & 0 & \cdots & I_B^m
\end{array} \right),
\end{equation}
and the interpolation and force summation operators would be extended as
\begin{equation}
\hat{E}^T = \left( \begin{array}{cccc}
E^T_1 & E^T_2 & \cdots & E^T_m
\end{array} \right), \ \ \ \  \hat{N}_B = \left( \begin{array}{cccc}
N_B^1 & 0 & \cdots & 0 \\ 
0 & N_B^2 & \cdots & 0 \\
\vdots & \vdots & \ddots & \vdots \\
0 & 0 & \cdots & N_B^m
\end{array} \right),
\end{equation}
while keeping the structure of block-LU decomposition exactly the same. When describing dense particle suspensions
one would need to complement the present method with appropriate collision model, such as the one employed by Kempe and Fr\"{o}hlich~\cite{kempe2012improved}.

\section{Validation} \label{sec:validation}

In this section, all validation cases are computed using the implicit coupling, as described in the previous section. First, we carry out temporal and spatial convergence tests for a simple FSI problem. We then validate the present method on a freely falling/rising cylinder, and on migration of a neutrally buoyant cylinder in a shear flow.

\subsection{Convergence}

\begin{figure}[t!]
  \centering
  \includegraphics[width=1.0\linewidth]{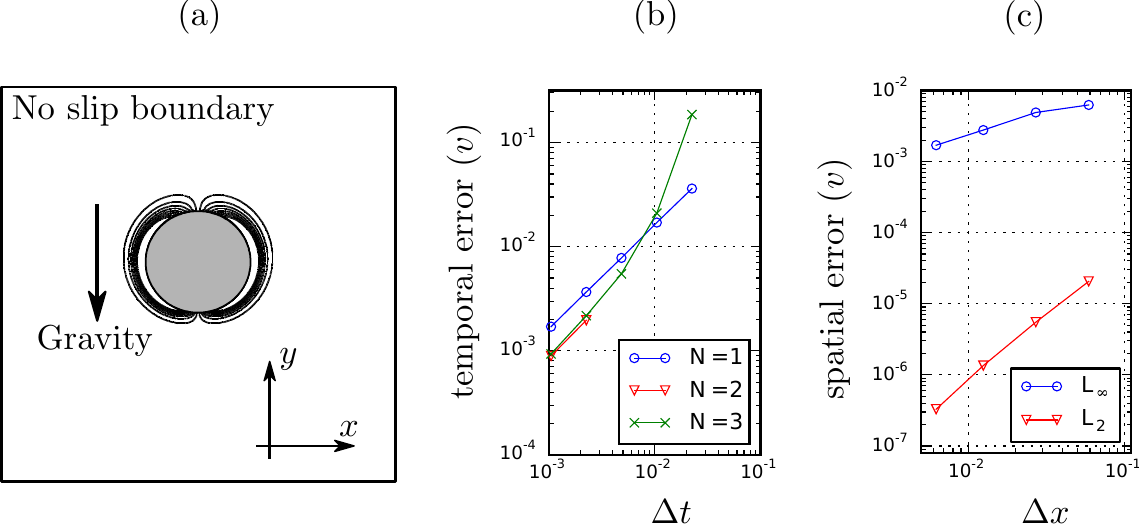}  
  \caption{A sketch of physical problem for convergence tests (a). The circular cylinder is placed in a box and free to fall in the gravitational field. All degrees of freedom are allowed for the circular cylinder (movement in $x$ and $y$ direction, as well as rotation). Black lines represent vorticity contours at the end of the simulation. The temporal error and spatial error is shown in frames (b) and (c), respectively. Temporal error is reported using up to three expansion terms ($N=3$)  of $A^{-1}$ (\ref{eq:proj-meth-BN-expansion}). }
\label{fig:conv-phys-skech}
\end{figure}

The convergence properties of the fluid solver and the IB method herein are reported by Perot~\cite{Perot_JCP_1993} and Taira and Colonius~\cite{Taira_JCP_2007}, respectively. We focus on the convergence when the fluid solver is coupled to rigid body dynamics. We select a simple two-dimensional problem -- a circular cylinder with diameter $D$ falling under the influence of gravity in a fluid with kinematic viscosity $\nu$ (see Fig.~\ref{fig:conv-phys-skech}a). The non-dimensional density of the cylinder is $\rho = 1.01$, the Reynolds number is $Re = D V_{\mathrm{term}} / \nu = 156$, where $V_{\mathrm{term}}$ is the terminal velocity, and the Gallileo number is $G = \sqrt{|\rho - 1| g D^3} / \nu = 138$.

To investigate the temporal convergence, we place the circular cylinder in the center of a square box with a width and a height of $8D$. We use a uniform mesh in the whole domain. For reference, we use a simulation with a moderate mesh spacing ($\Delta x = \Delta y = 0.01D$) and a small time step
($\Delta t = 10^{-4}$).
We carry out the simulation until
$t = 0.9$,
during which time the cylinder reaches the velocity
$V_{\mathrm{fall}} \approx 0.29 V_{\mathrm{term}}$.
We perform a set of simulations for a range of time steps 
$\Delta t \in \left[ 1.0 \times 10^{-3}, 2.25 \times 10^{-2} \right]$
on the same mesh. We define an infinity norm as the maximum of error $e_{i,j} = v^{ref}_{i,j} - v_{i,j}$, i.e. $L_{\infty} = \max |e_{i,j}|$.
The infinity norm for the temporal error is shown in Fig.~\ref{fig:conv-phys-skech}b, where it is observed that our method has
the same convergence rate as the original method for practical time steps, i.e. third-order in time (in current test $\Delta t > 10^{-2}$),
when a third-order approximation $\tilde{B}^3$ of the inverse of the Laplacian is used. In theory, the method is first-order in time due to 
the time-lagged interpolation and spreading operations.
The first-order error
starts to appear for smaller time steps (in current test $\Delta t < 10^{-2}$, see Fig.~\ref{fig:conv-phys-skech}b).
We found that the second-order approximation $\tilde{B}^2$ leads to unstable simulations for time steps larger than $\Delta t = 2.25 \cdot 10^{-3}$, which is similar to findings of Taira and Colonius~\cite{Taira_JCP_2007}; they showed that second-order approximation lacks positive-definiteness for higher time steps.
We observed that the order
of time derivative of integrals $\frac{d}{dt} \int_V \vec{u}\,\mathit{dV}$ and $\frac{d}{dt} \int_V \vec{r} \times \vec{u}\,\mathit{dV}$ plays a role in obtained convergence rate only for smaller time steps.

To investigate the spatial convergence, we reduce the box size to $2D \times 2D$. For reference, we use a simulation with a very fine mesh ($\Delta x = \Delta y = 0.00125D$) and a very small time step ($\Delta t = 10^{-5}$). We perform the reference simulation until $t = 0.01$, during which the cylinder reaches the velocity $V_{\mathrm{fall}} \approx 0.0035 V_{\mathrm{term}}$. We use a range of mesh spacings $\left[ 0.0063D, 0.059D \right]$, while keeping the time step constant. In order to compute the error $e_{i,j} = \tilde{v}^{ref}_{i,j} - v_{i,j}$, we interpolate the reference solution $v^{ref}_{i,j}$ on the coarser grid using third-order interpolation in space. We define a L2-norm as $L_2 = \sqrt{\sum e_{i,j}^2} / N_d$, where $N_d$ is the number of points in the error field. As shown in Fig.~\ref{fig:conv-phys-skech}c, the convergence rate is around $2$ in the L2-norm and around $1$ in the infinity norm, as observed by Taira and Colonius~\cite{Taira_JCP_2007} in their one-dimensional test case.

\subsection{Freely falling and rising cylinder}

There are a number of studies available for freely falling and rising bodies. For a review, we refer the reader to the work by Ern et al.~\cite{ern2012wake}. For physically relevant range of Reynolds numbers $\Rey$ (with density ratio $\rho$ close to unity), the falling cylinder problem has been investigated numerically by Namkoong et al.~\cite{Namkoong_JFM_2008}. They used an implicit coupling approach within a finite element method and adaptive body-fitted mesh with refined resolution in the cylinder wake.

\begin{figure}[t!]
  \centering
  \includegraphics[width=\linewidth]{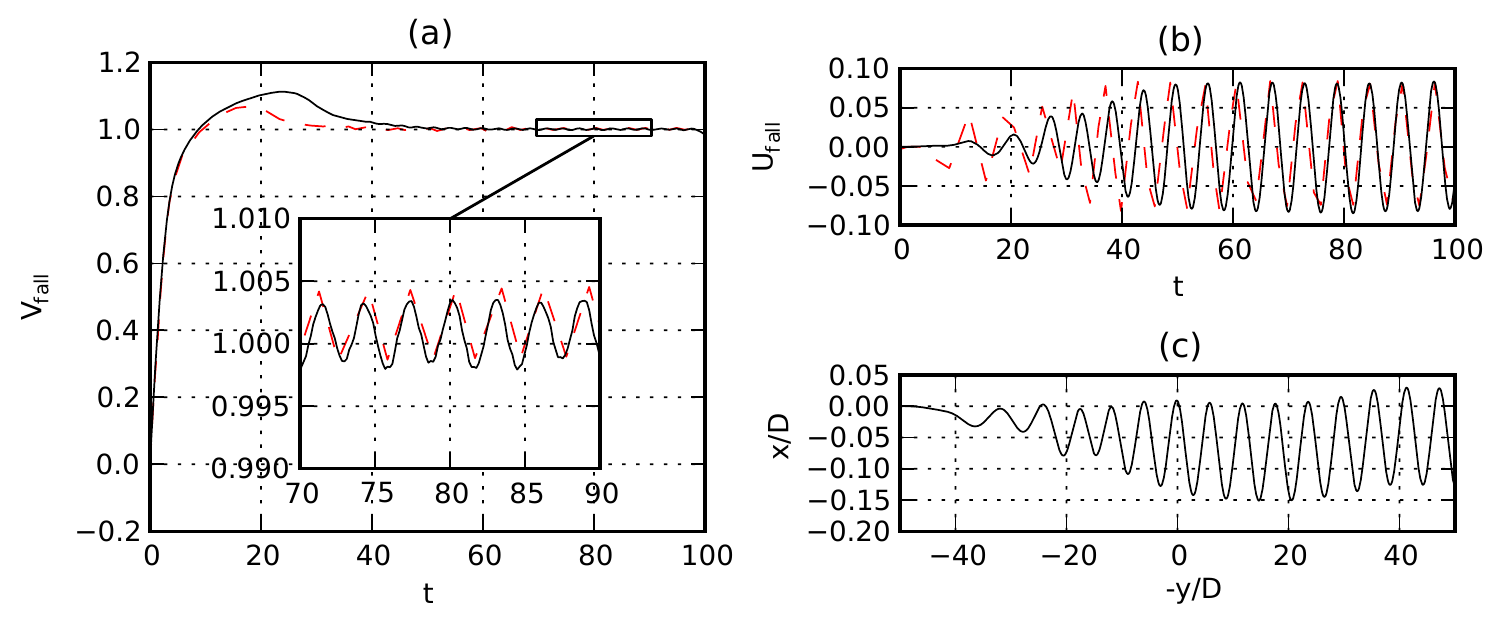} 
  \caption{Vertical (a) and horizontal (b) velocity, and trajectory (c) of freely falling circular cylinder, $\rho = 1.01$, and $\Rey = 156$:  ---, present results; \red{- - -}, results by~\cite{Namkoong_JFM_2008}.}
\label{fig:cyl-valid-fallSum}
\end{figure}

For validation purposes, a density ratio $\rho = 1.01$ and Reynolds number $\Rey = 156$ was selected. We carried out a simulation in a domain of size $ \left(x, y\right) \in [-5,5] D \times [ -50, 50 ] D$ using a uniform mesh ($\Delta x = \Delta y = 0.04D$) with linearly expanding mesh at the boundaries, and a CFL condition $V_{\mathrm{term}} \Delta t / \Delta x = 0.4$. The boundary condition at the exterior of the simulation domain is no-slip. The vertical velocity of the falling cylinder is compared to the results from~\cite{Namkoong_JFM_2008} in Fig.~\ref{fig:cyl-valid-fallSum}a. The agreement is satisfactory, despite the relative simplicity of the current simulation method. The difference in the transient regime, where the wake instability develops, can be explained by the difference of rates at which numerical error accumulates and breaks the symmetry of cylinder wake. The transverse velocity of the falling cylinder is shown in Fig.~\ref{fig:cyl-valid-fallSum}b, and the trajectory of the falling cylinder is shown in Fig.~\ref{fig:cyl-valid-fallSum}c.

\begin{figure}
  \centering
  \includegraphics[width=\linewidth]{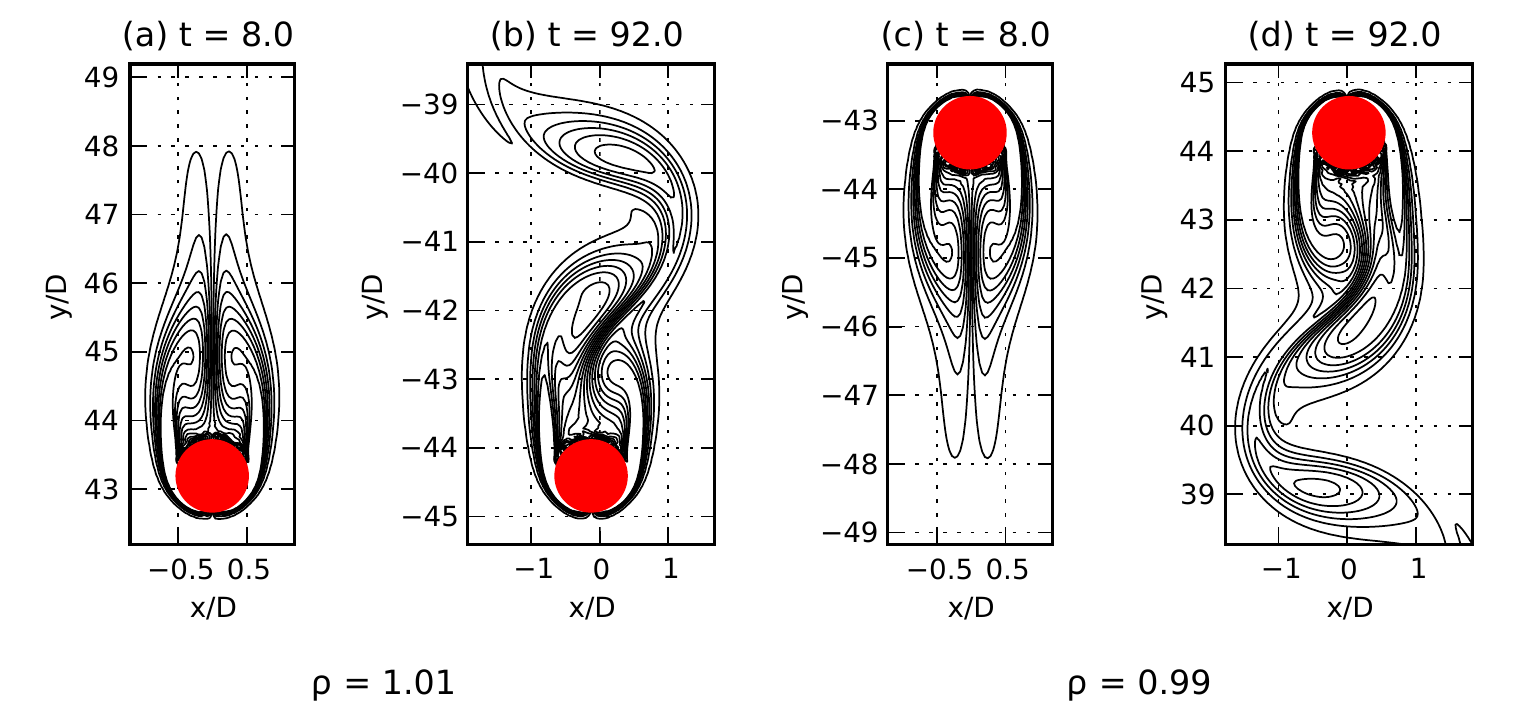} 
  \caption{Vorticity field contour lines around freely falling (a and b) and rising (c and d) circular cylinder at $\Rey = 156$. The vorticity levels from $\omega = -3.0$ to $3.0$ with step $\Delta \omega = 0.4$ are used. Flow field is reported at two different time values for both cases.}
\label{fig:cyl-valid-fallRiseVort}
\end{figure}

We present the vorticity field for freely falling and rising cylinders in Fig.~\ref{fig:cyl-valid-fallRiseVort} ($\omega_z = -3.0$ to $3.0$ with step $\Delta \omega_z = 0.4$). We observe that the flow field of falling (a and b) cylinder is symmetric to that around a rising (c and d) cylinder.
This is expected, since the Reynolds number for both configurations is the same and dimensionless densities ($\rho = 1.01$ and $0.99$) are very close to each other. Initially ($t = 8.0$), we see a symmetric vortex pair forming behind an accelerating cylinder. After the transient, flow perturbations have grown and the unstable symmetric solution has transitioned to the stable periodic vortex shedding state as seen in later times ($t = 92.0$). For the freely falling cylinder, we compare the Strouhal number $St = f_L D / V_{\mathrm{term}}$ -- where $f_L$ is frequency of lift force oscillations or vortex shedding --, drag coefficient value $C_D = F_{d} / (1/2 \rho_f V_{\mathrm{term}}^2)$ -- where $F_{d}$ is the drag force -- and amplitude of lift coefficient $C_L$ oscillations with values reported by Namkoong et al.~\cite{Namkoong_JFM_2008} in Tab.~\ref{tab:fall-rise-comp}, where we again observe satisfactory agreement. Although the flow fields are very similar, the freely rising cylinder has slightly higher Strouhal number, caused by the difference of the cylinder densities (Tab.~\ref{tab:fall-rise-comp}); same observation was reported in~\cite{Namkoong_JFM_2008}.

\begin{table}[ht!]
  \begin{center}
    \begin{tabular}{ | r | c | c | c | c | c | c | }
    \hline
      & \multicolumn{3}{|c|}{Falling} & \multicolumn{3}{|c|}{Rising} \\ \hline
      & $St$ & $C_D$ & $\max | C_L |$ & $St$ & $C_D$ & $\max | C_L |$ \\ \hline
    Present method & $0.17185$ & $1.29$ & $0.14$ & $0.17188$ & $1.29$ & $0.14$ \\ \hline
    Namkoong et al.~\cite{Namkoong_JFM_2008} & $0.16840$ & $1.23$ & $0.15$ & $0.16870$ & -- & -- \\ \hline
    \end{tabular}
  \end{center}
  \caption{Flow characteristics of freely falling and rising circular cylinder with density ratios $\rho = 1.01$ and $0.99$, respectively. For freely falling cylinder we report and compare values of Strouhal number, drag coefficient, and amplitude of lift coefficient. In work by Namkoong et al.~\cite{Namkoong_JFM_2008}, only the Strouhal number is reported for the freely rising body at $\Rey = 156$.} \label{tab:fall-rise-comp}
\end{table}

\subsection{Neutrally buoyant cylinder in shear flow}

\begin{figure}[t!]
  \centering
  \includegraphics[width=1.0\linewidth]{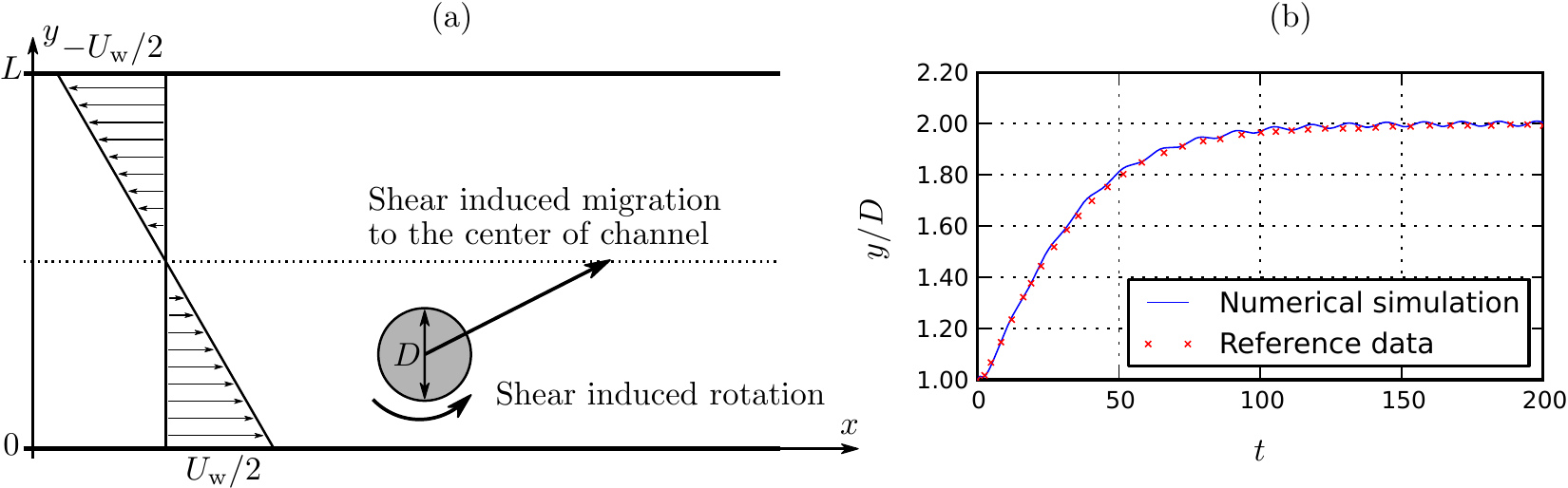} 
  \caption{Sketch of physical problem with significant rotation. Neutrally buoyant circular cylinder is placed in Couette flow (a). The cylinder migrates to the center of the channel. We show the vertical position of the cylinder with respect to time (b). We compare our results with findings of Feng et al.~\cite{feng1994direct} (red crosses).}
\label{fig:cyl-cylinder-Couette}
\end{figure}

In order to further validate the present numerical method, we consider the problem of neutrally buoyant circular cylinder of diameter $D$ migrating in a Couette flow (see Fig.~\ref{fig:cyl-cylinder-Couette}a). The channel height is $L = 4.0D$ with a uniform mesh in the $y$ direction. The channel width is $160D$ with coordinate $x \in \left[-80, 80\right]$ out of which the uniform mesh is used for $60D$ ($x \in \left[-30, 30\right]$). The uniform mesh spacing is $\Delta x = \Delta y = 0.04D$, and CFL number $U_\scr{w} \Delta t / \Delta x$ is set to $0.8$. The upper and lower walls move in the $x$-direction with velocities $- U_{\scr{w}} /2$ and $U_{\scr{w}} /2$, respectively, which gives a shear rate $\gamma = U_\scr{w}/L$. The Reynolds number is $Re = U_{\scr{w}} L / \nu = 40$, same as in the work of Feng et al.~\cite{feng1994direct}, which we use as a reference. They used a body-fitted mesh with a finite element-solver. Their configuration is also used by Feng and Michaelides~\cite{feng2004immersed}, Niu et al.~\cite{niu2006momentum}, and Bhalla et al.~\cite{Bhalla2013446} for testing the coupling between fluid and rigid body dynamics.

\begin{figure}[t!]
  \centering
  \includegraphics[width=1.0\linewidth]{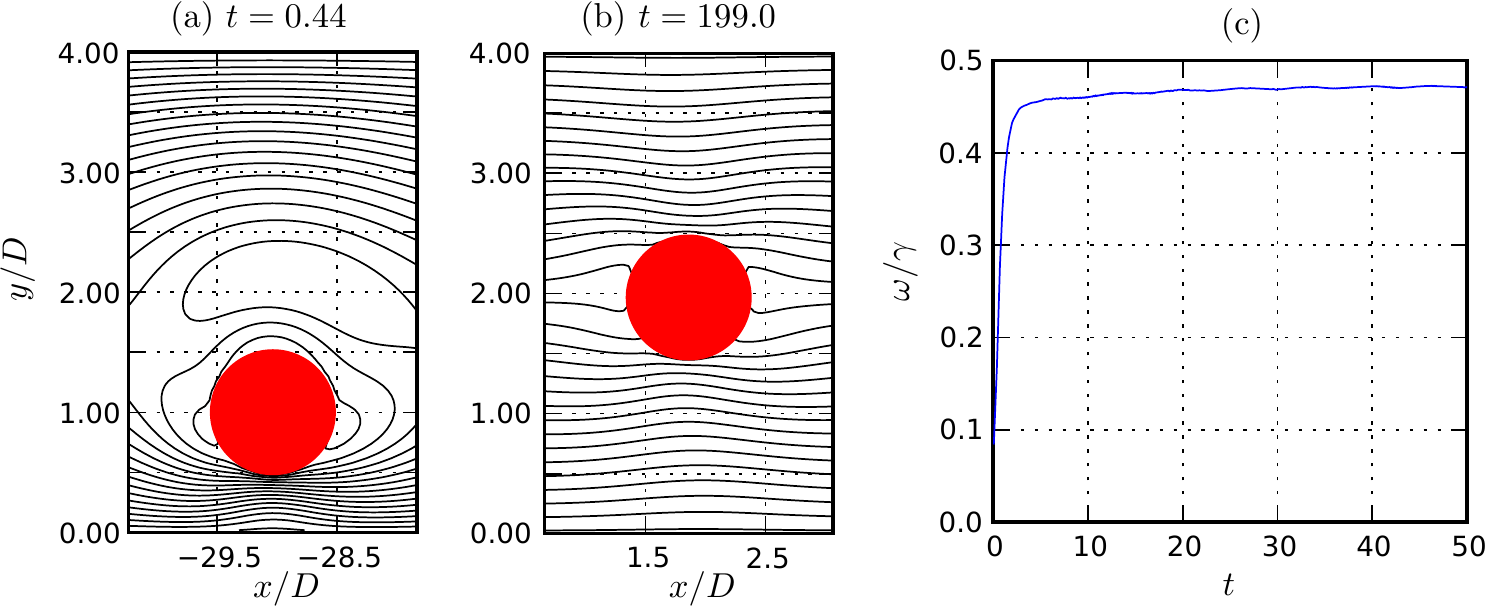} 
  \caption{Velocity field horizontal component $u$ contour lines around neutrally buoyant cylinder in a shear flow at $\Rey = 40$ (a and b). We show $32$ equally spaced horizontal velocity levels from $u = - U_\scr{w}/2$ to $U_\scr{w}/2$. Flow field is reported at two different time values. We also show the relaxation in time of normalized angular velocity $\omega/\gamma$.}
\label{fig:cyl-Couette-flow}
\end{figure}

We position the circular cylinder in the lower part of the channel ($y/L  = 0.25$) and release it with zero velocity and zero rotation. As observed in previous works~\cite{feng1994direct,feng2004immersed,niu2006momentum,Bhalla2013446}, the cylinder rotates and migrates from the release position to the center of the channel (see Fig.~\ref{fig:cyl-cylinder-Couette}b). The small oscillations in the trajectory (Fig.~\ref{fig:cyl-cylinder-Couette}b) are due to the fact that Lagrangian force in IB methods is dependent on the position of Lagrangian points relative to the fluid grid. Breugem~\cite{Breugem_JCP_2012} has noted a similar behavior observed in IB simulations and referred to it as ``grid locking''. Although we use the smooth 3-cell discrete delta function, the vertical migration is very slow compared to the rotation rate; and the error in vertical position associated to the ``grid locking'' can be seen. Consequently these oscillations illustrate the amplitude of error we have in the simulation.

We observe that rotation of the cylinder rapidly reaches a constant value, which is $47\%$ of the shear rate (see Fig.~\ref{fig:cyl-Couette-flow}c), as reported in prior studies~\cite{feng1994direct,feng2004immersed,niu2006momentum,Bhalla2013446}. To illustrate the flow field around the cylinder, we show $32$ equally spaced contour-lines of horizontal velocity $u$ from $-U_\scr{w}/2$ to $U_\scr{w}/2$ (Fig.~\ref{fig:cyl-Couette-flow}a,b). In case of a pure shear flow, the picture would consist of parallel lines only. When the cylinder is placed in the shear flow, initially -- cylinder has not reached the rotation defined by the shear rate -- there are significant distortions in the flow field (Fig.~\ref{fig:cyl-Couette-flow}a). When the cylinder has reached the terminal rotation rate, the modifications of flow field are minor;
the flow profile is compressed at the sides of the cylinder and flattened at the front and the back of the cylinder, which can be seen
in Fig.~\ref{fig:cyl-Couette-flow}b.

\section{Numerical stability of coupling between rigid body dynamics and fluid dynamics} \label{sec:stab-tests}

\begin{figure}[t!]
  \centerline{\includegraphics[width=\linewidth]{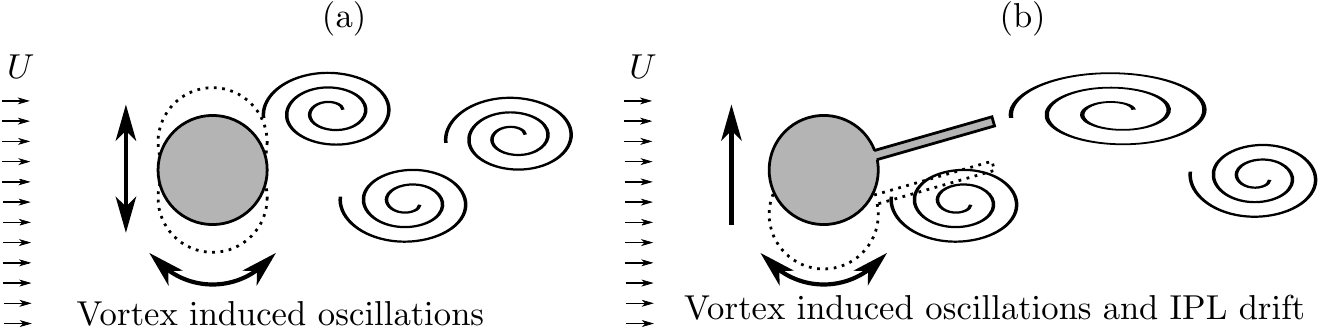}} 
  \caption{Two physical problems for numerical stability tests. Two degree of freedom motion of circular cylinder (a) and circular cylinder with splitter plate (b) in uniform free stream of Reynolds number $Re = 100$.}
\label{fig:stab-phys-skech}
\end{figure}

Let us examine the numerical stability of the current implicit method and compare it to an explicit coupling (see section~\ref{sec:expl-coup-sch}) between a rigid body and the fluid. As a model problem, let us consider a body placed in a uniform free stream, which can move in the cross stream direction and freely rotate (2 degrees of freedom). We test the numerical method using circular cylinder with and without a splitter-plate of length $L_{sp} = 1.0D$ clamped at the back of the cylinder. For the cylinder alone (see Fig.~\ref{fig:stab-phys-skech}a), we expect vortex induced vibrations of the cylinder, a classical fluid-structure interaction problem investigated thoroughly in the literature~\cite{sarpkaya2004critical,williamson2004vortex,Gabbai2005575}. For the cylinder with the splitter plate (see Fig.~\ref{fig:stab-phys-skech}b), we expect a drift caused by an inverted pendulum like (IPL) instability in addition to the VIV. As explained by L\={a}cis et al.~\cite{Lacis2014driftSymmetry}, the
body orientation, when the splitter plate
is parallel to the incoming free stream is always an equilibrium solution to the fluid-structure interaction problem. However, when the plate is sufficiently long, this solution
becomes unstable in a manner similar to how an inverted pendulum becomes unstable under gravity.
When this instability is triggered, the body turns until it reaches a new equilibrium turn angle, and it steadily drifts in the direction, in which the splitter plate has turned.
We choose $Re = U D / \nu = 100$ based on the free stream velocity $U$, the cylinder diameter $D$ and the kinematic viscosity of fluid $\nu$. There is no spring $k = 0$ and no damper $b = 0$ associated with the body.

\begin{figure}[t!]
  \vspace*{-10pt}
  \centering
  \includegraphics[width=0.495\linewidth]{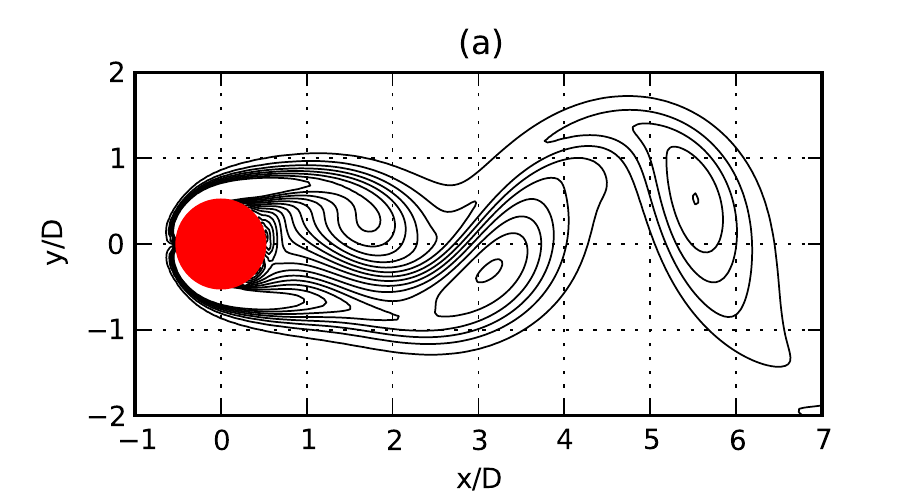}
  \hfill
  \includegraphics[width=0.495\linewidth]{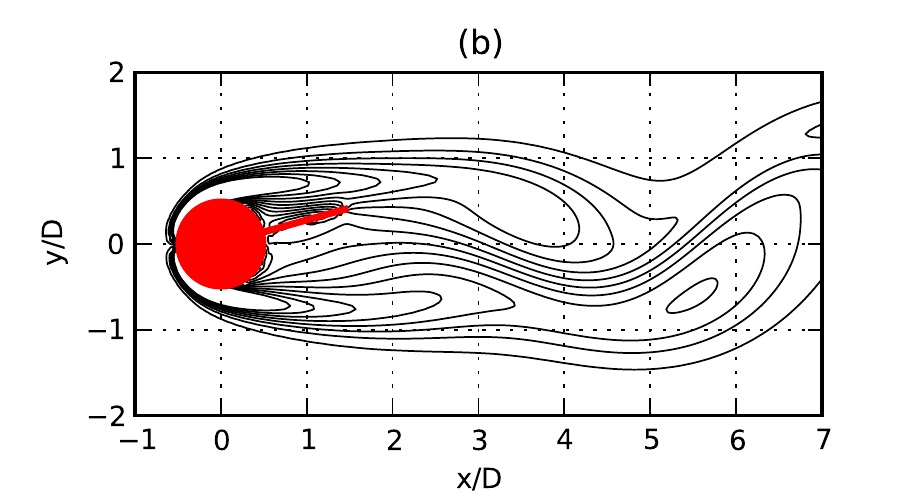} \\
  \vspace*{-10pt}
  \caption{Initial flow conditions for numerical stability tests. Flow around a circular cylinder (a) and flow around circular cylinder with splitter plate (b) in uniform free stream of Reynolds number $Re = 100$ after transient has decayed. The vorticity levels from $\omega = -3.0$ to $\omega = 3.0$ using step $\Delta \omega = 0.4$ are shown.}
\label{fig:stab-phys-initflow}
\end{figure}

We place the cylinder in the center of the computational domain of size $\left(x, y\right) \in [-15,45] D \times [ -20, 20 ] D$.
The mesh is uniform in the region $\left(x, y\right) \in [-1,3] D \times [ -2, 2 ] D$ 
for the flow around cylinder and
$ \left(x, y\right) \in [-1,3] D \times [ -1, 9 ] D$ 
for the flow around the cylinder with the splitter plate. The uniform grid spacing is $\Delta x = \Delta y = 0.04D$ and the CFL number is $U \Delta t / \Delta x = 0.4$.
The initial conditions
shown in Fig.~\ref{fig:stab-phys-initflow}a and b have been obtained by performing simulations, in which we constrain all degrees of freedom for the body (fix the body) and let the flow evolve around it for $100$ time units.

\begin{figure}[t!]
  \centering
  \includegraphics[width=1.0\linewidth]{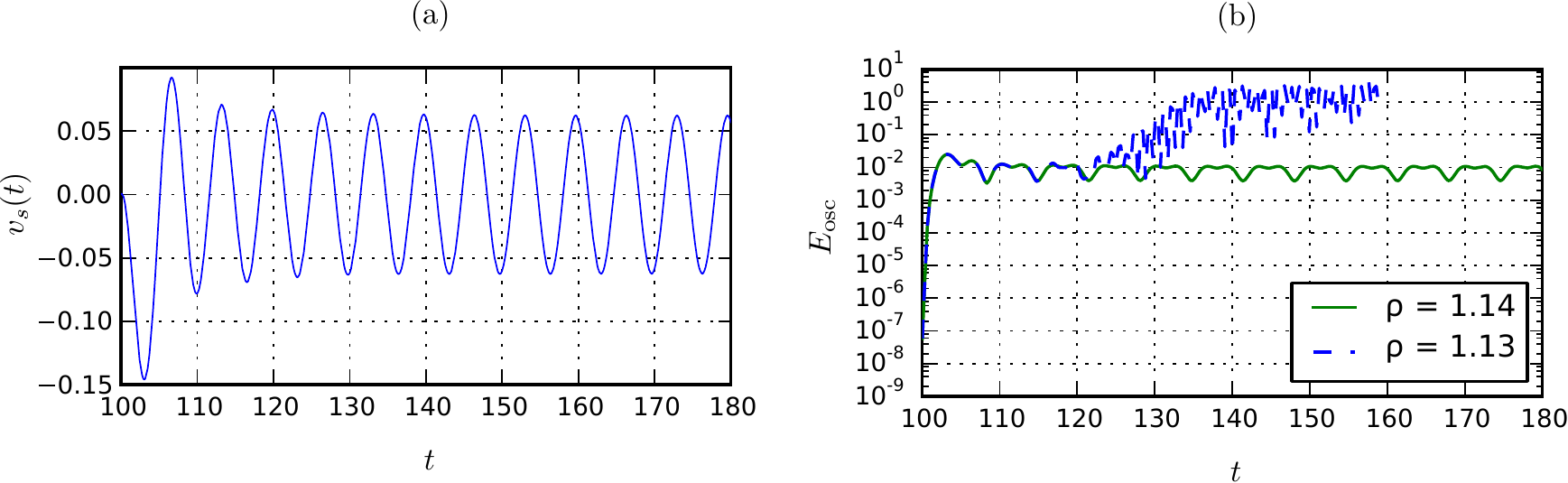} 
  \caption{Results of cylinder motion simulation. We show vertical velocity $v_s \left(t\right)$ for cylinder with density ratio $\rho = 1.14$ (a). We also show the spring oscillator energy $E_{\mbox{osc}}$ for cylinder with density ratios $\rho = 1.14$ and $1.13$ (b). Here we report simulations with inverse matrix $B_N$ approximation order $N=3$.}
\label{fig:stab-Euler-cyl}
\end{figure}

\begin{figure}[t!]
  \centering
  \includegraphics[width=0.7\linewidth]{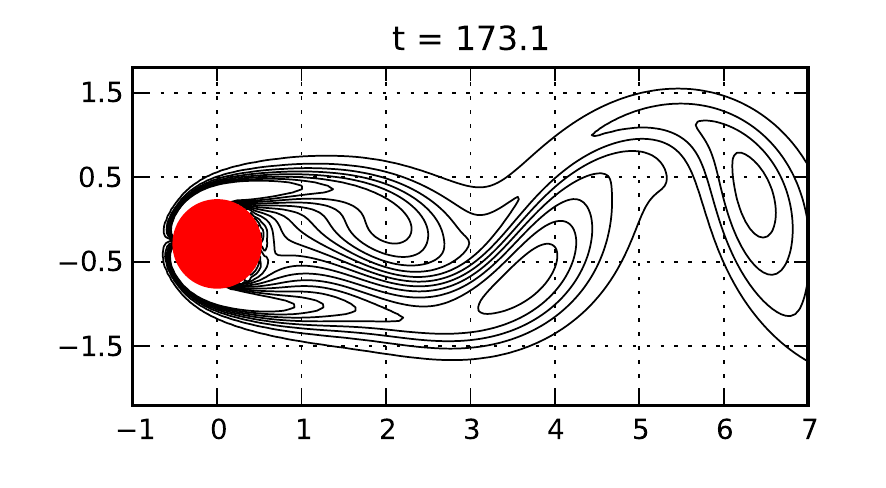}
  \includegraphics[width=0.7\linewidth]{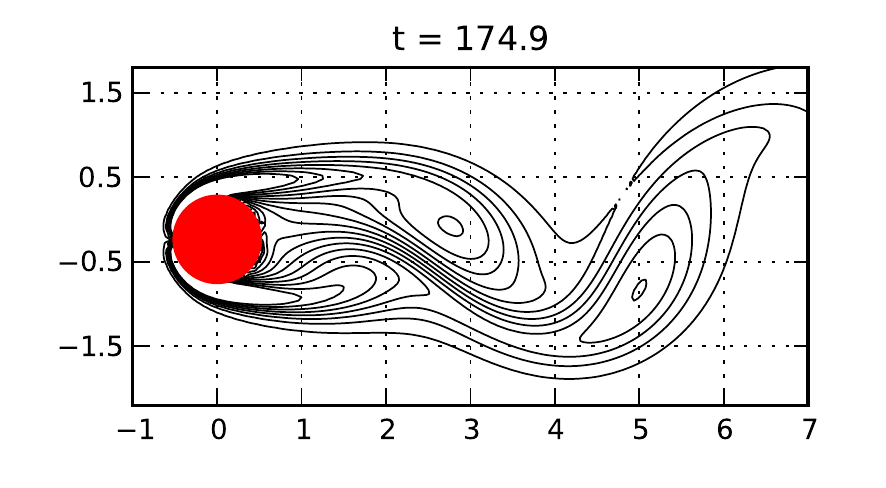}
  \includegraphics[width=0.7\linewidth]{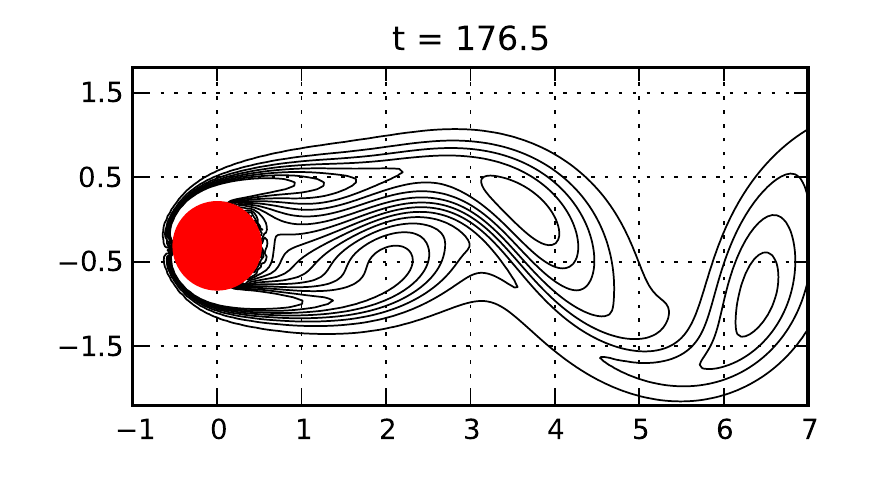}
  \caption{The VIV dynamics of a circular cylinder as vortex is being formed and shed from the lower part of the cylinder. At the initial stage of the vortex formation ($t = 173.1$), the cylinder is moving upwards and in same time forced downwards by smaller pressure field of the arising vortex. At the intermediate stage ($t = 174.9$), the cylinder has reached highest vertical position and is at still. Finally, the cylinder is accelerating downwards, until vortex is shed ($t = 176.5$) and pressure force change direction due to another vortex being formed at the upper side of the cylinder. $\Rey = 100$, $\rho = 1.14$, $k = 0$, and $b = 0$.}
\label{fig:stab-phys-viv-desc}
\end{figure}

In order to determine whether Euler's forward method (see section \ref{sec:expl-coup-sch}) for coupling the rigid body and the fluid is stable for a given density ratio $\rho$, we
carry out the simulation from $u_s = v_s = \omega_s = 0$ (stationary body) until the transient behavior has decayed and there are a number of periodic oscillations visible. Fig.~\ref{fig:stab-Euler-cyl}a shows the vertical velocity $v_s \left(t\right)$ for stable simulation of cylinder with $\rho = 1.14$.
We observe that after a short transient, the cylinder has reached a periodic transverse oscillatory state. Here, the so called VIV phenomenon~\cite{sarpkaya2004critical,williamson2004vortex,Gabbai2005575} has been reproduced -- vortex formation at the lower side of the cylinder is shown in Fig.~\ref{fig:stab-phys-viv-desc}.

When the density ratio is reduced to $\rho = 1.13$,
we observe that
after some time (approximately $25$ time units) the scheme is unstable.
The instability is illustrated in Fig.~\ref{fig:stab-Euler-cyl}b by the oscillation energy
\begin{equation}
E_{\mbox{osc}} = v_s^2 + k_{\scr{osc}} y^2,
\end{equation}
where $k_{\scr{osc}}$ is an effective spring constant, and $y$ is the position of the body. The oscillation energy $E_\scr{osc}$ is based on kinetic energy of the cylinder and potential energy of a virtual spring; we neglect damping and forcing terms (for a complete VIV model see~\cite{williamson2004vortex}). We set the effective spring constant to value $k_{\scr{osc}} = \max\limits_{t>120} |v_s|^2 / \max\limits_{t>120} |y|^2$, such that the oscillation energy $E_\scr{osc}$ is the same, when the transverse velocity $v_s$ has maximal and zero values in the steady oscillation regime ($t > 120$).
The critical density ratio $\rho_c$ is defined as the lowest value, for which we found the simulation to be stable, i.e. $\rho_c = 1.14$.

We carry out stability tests
also for the remaining coupling algorithms, i.e. Heun's method (section~\ref{sec:expl-coup-sch}) and implicit method (section~\ref{sec:imp-coupl-desc}), for the cylinder with and without the splitter plate. All stability tests are performed over $80$ time units using an inverse Laplacian matrix $\tilde{B}^N$ with approximation order $N = 1$ to save computational time.
A summary of the obtained critical density ratios $\rho_c$ is listed in Tab.~\ref{tab:stab-summary}. 
It is observed that $\rho_c$ does not depend on the time step (in line with findings of~\cite{Borazjani20087587}, where they report no influence of time step on the numerical stability of vortex induced vibrations) and approximation order $N$.

\begin{table}[t!]
  \begin{center}
    \begin{tabular}{ | r | c | c | c | }
    \hline
      & Euler's method & Heun's method & Implicit \\ \hline
    Cylinder & $1.14$ & $0.37$ & $10^{-4}$ \\ \hline
    Cylinder with splitter plate & $8.91$ & $4.64$ & $10^{-4}$ \\ \hline
    \end{tabular}
  \end{center}
  \caption{The lowest density ratio $\rho_c$, which we found the Euler's method (\ref{eq:rbd-Euler1}--\ref{eq:rbd-Euler3}), Heun's method (\ref{eq:rbd-Heun1}--\ref{eq:rbd-Heun6}) and present method (\ref{eq:def-impl1}--\ref{eq:def-impl3}) to be stable. In these tests we evaluate the motion of fluid inside particle explicitly. We tested cylinder with and without a splitter plate of length $L_{sp} = 1.0D$.} \label{tab:stab-summary}
\end{table}

We can observe from Tab.~\ref{tab:stab-summary} that adding a splitter plate behind a cylinder increases the value of critical density ratio $\rho_c$ for the explicit coupling by nearly an order of magnitude. It has been previously reported that
the ratio between the fluid forcing and the inertia
of the body is a parameter, which enters stability criterion for explicit coupling methods (see~\cite{Conca1997387,Borazjani20087587}).
Borazjani et al.~\cite{Borazjani20087587} analyzed both weak and strong iterative FSI coupling, where they show that in the case of a low object mass, both numerical schemes can yield unstable results.
The rapid increase of $\rho_c$, by adding a splitter plate, for explicit coupling methods (Tab.~\ref{tab:stab-summary}) confirms that 
the ratio between the fluid forcing and the inertia of the body is a stability parameter. By introduction of splitter plate, we have significantly increased the resulting torque on the body, and in order to preserve the stability of algorithm, the inertia of the body (density) had to be increased as well.
However, the implicit coupling is stable down to $\rho = 10^{-4}$ for the cylinder with and without the splitter plate.
It suggests that the ratio between fluid forcing and body inertia is no longer a parameter for stability.
Our approach does not suffer from the instability of SC observed by Borazjani et al.~\cite{Borazjani20087587}, because
the convergence is guaranteed by the positive-definiteness of the algebraic equations in our method.

We note that the proposed implicit coupling becomes unstable when density ratio exactly becomes $\rho = 0.00$, because the matrix $\tilde{B}^N$ (\ref{eq:matrix-tildeBn}) is then singular.
An alternative approach to solve this issue using a block-LU decomposition is discussed in the Appendix (Stability of the implicit coupling for massless particles).

\section{Significance of fictitious fluid motion inside particle} \label{sec:fict-fluid}

Uhlman~\cite{Uhlmann_JCP_2005} assumed that fluid inside particle follows rigid body motion.
Breugem~\cite{Breugem_JCP_2012} argued that the fluid inside a particle does not follow the rigid body motion due to an error in fluid velocity near the surface of the body and therefore approach suggested by Kempe and Fr\"{o}hlich~\cite{kempe2012improved} should be implemented. Since in our method the boundary condition for interpolated fluid velocity on the body surface is imposed up to machine precision, we could expect the assumption of Uhlman to be valid. 
Under this assumption, the linear and angular fluid acceleration terms in equations (\ref{eq:Newt1-ibm}--\ref{eq:Newt2-ibm}) can be simplified as
\begin{equation}
\frac{d}{dt} \int_V \vec{u}\,\mathit{dV} = V_s \frac{d\vec{u}_s}{d t},\qquad \frac{d}{dt} \int_V \vec{r} \times \vec{u}\,\mathit{dV} = I_s \frac{d\omega_s}{d t},
\end{equation}
which results in Newton's equations of motion in the following form
\begin{align}
\left( \rho - 1 \right) V_s \frac{d\vec{u}_s}{d t} & = - \oint_{S} \vec{F}\,\mathit{dS} + V_s \left( \rho - 1 \right) g \hat{e}_g, \label{eq:Newt1-ibm-Uhlm} & \\
\left( \rho - 1 \right) I_s \frac{d\omega_s}{d t} & = - \oint_{S} \vec{\tilde{\mathcal{L}}} \times \vec{F}\,\mathit{dS}. \label{eq:Newt2-ibm-Uhlm} &
\end{align}

\begin{figure}[t!]
  \centering
  \includegraphics[width=1.0\linewidth]{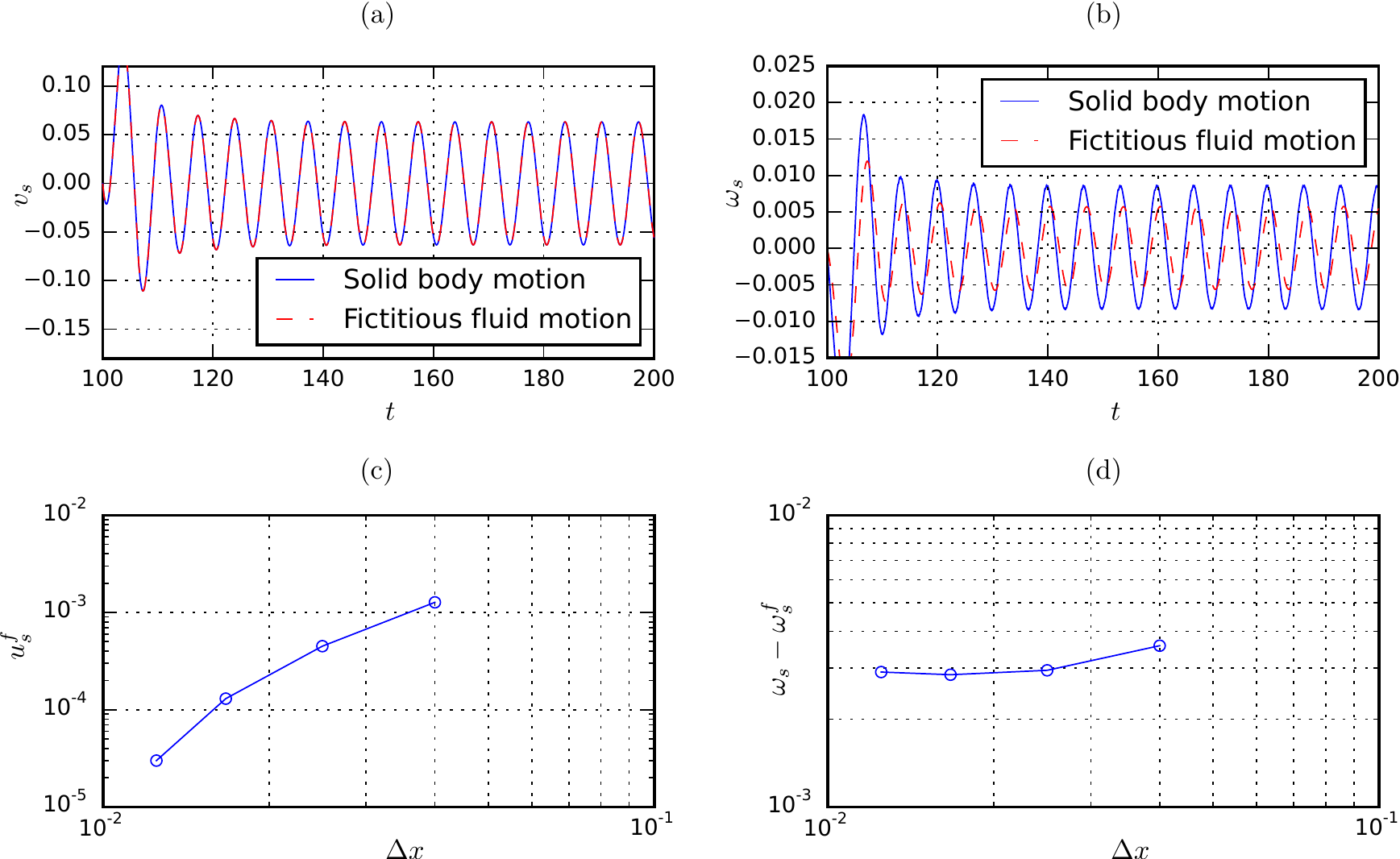}
  \caption{Results of cylinder motion simulation. We show vertical velocity $v_s \left(t\right)$ (a) and angular velocity $\omega_s \left(t\right)$ (b) for cylinder with density ratio $\rho = 1.01$, and compare results from solid body solver and integral of fluid inside particle (fictitious fluid), mesh resolution $\Delta x = 0.0125D$. In addition, we present convergence curves of horizontal fictitious fluid velocity $u_s^f$ (c), which should follow the imposed zero velocity, and also difference between peaks of solid body angular velocity $\omega_s$ and fictitious fluid angular velocity $\omega_s^f$ (d). }
\label{fig:expl-int-comp}
\end{figure}

To validate the rigid body assumption, we use the same test problem from section~\ref{sec:stab-tests}, which is vortex-induced-vibrations (VIV) of a circular cylinder with and without a splitter plate. We solve rigid body equation in Uhlman's form (\ref{eq:Newt1-ibm-Uhlm}--\ref{eq:Newt2-ibm-Uhlm}) and compare the motion of the cylinder ($\vec{u}_s$ and $\omega_s$) with
\begin{equation}
\vec{u}_s^f = \frac{1}{V_s} \int_V \vec{u}\,\mathit{dV},\ \ \omega_s^f = \frac{1}{I_s} \int_V \vec{r} \times \vec{u}\,\mathit{dV}.
\end{equation}
We carry out the computation using the same computation parameters as in section~\ref{sec:stab-tests}, while changing the grid spacing.
The computed vertical velocity and angular velocity are shown in Fig.~\ref{fig:expl-int-comp}a,b
for grid spacing $\Delta x = 0.0125D$.
While the fluid flow inside the particle follow the translational velocity of particle very well (Fig.~\ref{fig:expl-int-comp}a), the angular velocity is smaller in amplitude and lags in phase compared to the motion of solid body (Fig.~\ref{fig:expl-int-comp}b).
Translational velocity of the fluid particles inside the body is enforced by the translational velocity of the boundary through pressure field and incompressibility. Since we also enforce incompressibility up to machine precision, one may expect that fluid motion inside the particle follows the rigid body motion quite well. The fluid rotation on the other hand is enforced by viscous friction, therefore one cannot expect an immediate reaction due to changes in velocity value at the boundary of the cylinder.
In order to make sure that the discrepancy is not caused by spatial discretization error, we carry out the same test for other grid spacings $\Delta x = 0.0167D, 0.025D$ and $0.04D$. We then compute the difference between the expected solid body motion and the actual motion of fictitious fluid inside. Results for translational velocity component $u_s$ and angular velocity $\omega_s$ are shown in Fig.~\ref{fig:expl-int-comp}c,d, respectively. We observe that while the translational velocity of the fictitious fluid continuously converges to the expected value, the difference in angular velocity converges to some non-zero value. This confirms that the discrepancy between the fictitious fluid motion and the solid body motion does not come from spatial discretization error and is
due to the treatment of the interior (existence of fictitious fluid) of the solid body.

We note that there is an alternative approach to deal with the fluid inside the particle. That is to use IB forcing all over the volume of the particle to enforce the rigid body motion inside the particle as employed by Glowinski et al.~\cite{glowinski1999distributed}. Theoretically another possible approach is to explicitly evaluate the fluid stress tensor $\ten{\tau}$ and integrate it over the surface of the particle. However, it is highly non-trivial to rigorously define the boundary of the particle on the Eulerian grid with discrete delta functions spanning over multiple fluid cells.

\begin{table}[t!]
  \begin{center}
    \begin{tabular}{ | r | c | c | c | c | }
    \hline
      & Euler's method & Heun's method & Implicit & Implicit 2  \\ \hline
    Cylinder & $2.12$ & $1.58$ & $1.00$ &  $0.28$  \\ \hline
    Cylinder w.s.p. & $9.89$ & $5.59$ & $1.00$ & $0.00$ \\ \hline
    \end{tabular}
  \end{center}
  \caption{Critical density ratio $\rho_c$, at which the selected coupling between Newtons equations of motion and the Navier-Stokes equations becomes unstable. In these tests we have assumed that fluid inside particle exactly follows rigid body dynamics (Uhlman's assumption). In ``implicit 2'' method the fictitious fluid description using Uhlman's assumption is made explicit and moved to right hand side of governing equation system. We tested cylinder alone and cylinder with splitter plate of length $L_{sp} = 1.0D$.} \label{tab:stab-summary-implInt}
\end{table}

We also found that the assumption of rigid body motion inside the body has an effect on the numerical stability.
Using this assumption, there are two ways to formulate the equations. The first and most commonly used approach is to combine the inertia of fictitious fluid with the inertia of solid body itself, changing the prediction step (\ref{eq:NewtMat-proj1}) to
\begin{equation}
\left( \begin{array}{cc}
A & 0  \\
0 & I^U_B
\end{array} \right)  \left( \begin{array}{c}
q^{*} \\
u^{*}_B
\end{array} \right) = \left( \begin{array}{c}
r^{n} + bc_1 \\
(r^U_B)^{n}
\end{array} \right), \label{eq:proj-step-1-Uhlman}
\end{equation}
where the diagonal matrix of inertia (\ref{eq:diag-mat-inert}) is modified to become
\begin{equation}
I^U_B = \frac{\rho - 1}{\Delta t} \left( \begin{array}{ccc} V_s & 0 & 0 \\
0 & V_s  & 0 \\
0 & 0 & I_s \end{array} \right), \label{eq:dig-mat-inert-Uhlm}
\end{equation}
and there is no contribution from motion of fictitious fluid in the right-hand side term $(r^U_B)^{n}$. The second approach would be to leave the diagonal matrix of inertia unmodified, but change the right-hand side in the prediction step to
\begin{equation}
\left( \begin{array}{cc}
A & 0  \\
0 & I_B
\end{array} \right)  \left( \begin{array}{c}
q^{*} \\
u^{*}_B
\end{array} \right) = \left( \begin{array}{c}
r^{n} + bc_1 \\
r^{n}_B \left( dQ_B^U \right)
\end{array} \right), \label{eq:proj-step-1-Uhlman2}
\end{equation}
where the right-hand side depends on fictitious fluid motion the same way as in the proposed formulation (\ref{eq:NewtMat-proj1}), and the integral is substituted with explicit time derivative of solid body motion
\begin{equation}
dQ^U_B = \left( \begin{array}{ccc} \frac{u_s^{n}-u_s^{n-1}}{\Delta t} & \frac{v_s^{n}-v_s^{n-1}}{\Delta t} & \frac{\omega_s^{n}-\omega_s^{n-1}}{\Delta t} \end{array} \right)^T.
\end{equation}
We denote the approach with modified diagonal matrix of inertia (\ref{eq:proj-step-1-Uhlman}) as the ``implicit method with Uhlman's assumption'', and the approach with original diagonal matrix of inertia (\ref{eq:proj-step-1-Uhlman2}) as the ``implicit method 2 with Uhlman's assumption''.
We repeated tests from section~\ref{sec:stab-tests} of critical density ratio $\rho_c$ using Newton's equations of motion with Uhlman's assumption for all explicit and implicit couplings, results are shown in Tab.~\ref{tab:stab-summary-implInt}. Using this assumption, the present method with modified matrix of inertia becomes unstable for bodies with density ratio of unity.

The reason is that matrix $I^U_B$ (\ref{eq:dig-mat-inert-Uhlm}) is singular for density ratio $\rho = 1$, similarly as matrix $I_B$ (\ref{eq:diag-mat-inert}) is singular, when $\rho = 0$.
The present method using Uhlman's assumption but keeping the fictitious fluid inertia on the right-hand side of equations achieves much better stability properties compared to method with modified matrix of inertia for cylinder with splitter plate, with only slight improvement in the cylinder case. It is an interesting observation, since for the explicit coupling the stability properties for cylinder with splitter plate is worse compared to cylinder alone. 
Further examination is needed to uncover what causes
this change of behavior.

Although in simulations of the VIV problem with and without Uhlman's assumption we did not notice any accuracy problems, our findings suggest that if the rotation of the solid body is significant, the fluid motion inside particle must be taken into account, regardless if velocity of solid body boundary is imposed accurately or not. Even if rotation is not significant, there can be an effect on stability of the numerical scheme, as seen by comparing Tab.~\ref{tab:stab-summary-implInt} and Tab.~\ref{tab:stab-summary}.
We have also noticed that some of the improvement with respect to numerical stability can be achieved using Uhlman's assumption and separating the fictitious fluid inertia from the inertia of the solid body itself, as observed by comparing ``Implicit'' and ``Implicit 2'' methods in Tab.~\ref{tab:stab-summary-implInt}. However, the best stability properties have been achieved using the explicit integration of fictitious fluid inside the solid body.

\section{Conclusions} \label{sec:conclusions}

In the current work, we have extended the IB projection method~\cite{Taira_JCP_2007} to fluid-structure interaction problems with explicit and implicit coupling to rigid body dynamics. We showed that the second-order in space and third-order in time accuracy for practical time steps is preserved from the original method.
In addition, there is no added computational cost in the modified pressure Poisson step (size of the algebraic system is the same), while prediction and projection steps are complemented by only few rows and columns. 
The developed method has shown excellent stability properties for wide range of density ratios for both cylindrical and non-cylindrical bodies.
The influence on the accuracy and stability of the fictitious fluid flow inside the body has been examined in detail. In order to treat inertia of a solid body accurately within IB methods, the effect of the fictitious fluid flow has to be directly taken into account. For the stability of the coupled system it is necessary to(i) separate the inertia of fictitious fluid from the inertia of the solid body and (ii) take into account the fictitious fluid flow $dQ_B$ (\ref{eq:fict-fluid-int-all}).
As a final remark, we note that the block-LU decomposition has once again proved to be a powerful analysis tool for deriving algorithms to efficiently approximate solutions of fluid-structure interaction problems.

\section*{Acknowledgments}

U.L and S.B. are grateful for financial support from the Swedish Research Council (VR-2010-3910). U.L. acknowledges his colleagues at KTH Mechanics for sharing computational power of their workstations during heavy testing of the developed method. K.T. thanks support from the US Army Research Office (Grant number W911NF-13-1-0146).

\section*{Appendix. Stability of the implicit coupling for massless particles}

The method outlined in current work is unstable for the density value exactly $\rho = 0$ as seen in Tab.~\ref{tab:stab-summary}. This instability arises form the requirement to compute inverse of inertia matrix $I_B$ as part of $\tilde{B}^N$, which is singular when $\rho = 0$. If we rewrite equations for rigid body dynamics (\ref{eq:Newt1-ibm} -- \ref{eq:Newt2-ibm}) for $\rho = 0$, we obtain
\begin{align}
0 & = - \oint_{S} \vec{F}\,\mathit{dS} + \frac{d}{dt} \int_V \vec{u}\,\mathit{dV} - V_s g \hat{e}_g, \label{eq:Newt1-ibm-zero} & \\
0 & = - \oint_{S} \vec{\tilde{\mathcal{L}}} \times \vec{F}\,\mathit{dS} + \frac{d}{dt} \int_V \vec{r} \times \vec{u}\,\mathit{dV}, \label{eq:Newt2-ibm-zero} &
\end{align}
which is a dynamic condition for the rigid body motion without explicit acceleration term for the solid body.
The outlined block-LU decomposition cannot satisfy this condition. In order to overcome this limitation, one can rearrange the discrete equations. The algebraic form (\ref{eq:full-mat-Newton}) can be rearranged to
\begin{equation}
\left( \begin{array}{cccc}
A & G & E_n^T & 0 \\
G^T & 0 & 0 & 0 \\
E_n  & 0 & 0 & (N^n_B)^T	\\
0 & 0 & N^n_B  & I_B
\end{array} \right) \left( \begin{array}{c}
q^{n+1} \\
\phi^{n+1/2} \\
\tilde{f}^{n+1/2} \\
u_B^{n+1}
\end{array} \right) = \left( \begin{array}{c}
r^n \\
0 \\
\Delta u_B^{n+1} \\
r^{n}_B
\end{array} \right) + \left( \begin{array}{c}
bc_1 \\
- bc_2 \\
0 \\
0
\end{array} \right), \label{eq:full-mat-Newton-rearr}
\end{equation}
where Newton's equations of motion have been shifted to the lower part of the matrix and non-zero density ratio $\rho \neq 0$ is used for generality. As the matrix now has non-zero elements in the lower right matrix block, the block-LU decomposition now yields
\begin{equation}
\left( \begin{array}{cc}
A & \tilde{Q} \\
\tilde{Q}^T & C
\end{array} \right) = \left( \begin{array}{cc}
A & 0 \\
\tilde{Q}^T & C - \tilde{Q}^T B^N \tilde{Q}
\end{array} \right) \left( \begin{array}{cc}
I & B^N \tilde{Q} \\
0 & I
\end{array} \right). \label{eq:general-decomp}
\end{equation}
We have defined the block matrices as
\begin{equation}
\tilde{Q}^T = \left( \begin{array}{c}
G^T \\
E_n  \\
0
\end{array} \right), \ \ C = \left( \begin{array}{ccc}
0 & 0 & 0 \\
0 & 0 & (N^n_B)^T	\\
0 & N^n_B  & I_B
\end{array} \right),
\end{equation}
which leads to following modified Poisson equation
\begin{equation}
\left( \begin{array}{ccc}
G^T B^N G & G^T B^N E_n^T & 0 \\
E_n B^N G & E_n B^N E_n^T & -(N^n_B)^T \\
0 & -N^n_B & -I_B
\end{array} \right) \left( \begin{array}{c}
\phi^{n+1} \\ \tilde{f}^{n+1} \\ u_B^{n+1}
\end{array} \right) = \left( \begin{array}{c}
G^T \\ E \\ 0
\end{array} \right) q^* - \left( \begin{array}{c}
-bc_2 \\ \Delta u_B^{n+1} \\ r_B^n
\end{array} \right).
\end{equation}
The velocity is solved in the same matrix as IB force, therefore one can find a force, which satisfies dynamic conditions (\ref{eq:Newt1-ibm-zero} -- \ref{eq:Newt2-ibm-zero}). Nevertheless, this reordering results in a Poisson system, which is not positive definite.

We also note that reordered approach could potentially be useful in other cases, for example, to implicitly incorporate collision models. While the Poisson equation is not positive-definite, the prediction step for fluid velocity is still positive-definite and can be solved efficiently, regardless of what equations are solved in Poisson step.

\section*{Appendix. Designing a parallel Poisson solver by using the block-LU decomposition}

The most common designs of parallel algorithms rely on domain partitioning either in space, or in time~\cite{ferziger2002computational}. We show that block-LU decomposition can be used to design a parallel algorithm for Poisson solver, which does not rely on domain decomposition. We apply the block-LU decomposition on the Poisson equation matrix (\ref{eq:NewtMat-proj2}) and use the following notation for the matrix blocks
\begin{align}
M_{GBG} & = G^T B^N G,\ \ & M_{GBE} = G^T B^N E_n^T, \\
M_{EBE}  &= E_n B^N E_n^T + N_B^T I_B^{-1} N_B,
\end{align}
for which the block-LU decomposition of the modified pressure Poisson equation matrix (\ref{eq:NewtMat-proj2}) becomes
\begin{align}
& \left( \begin{array}{cc}
M_{GBG} & M_{GBE} \\
M_{GBE}^T  & M_{EBE}
\end{array} \right) \label{eq:second-blockLU} \\
& = \left( \begin{array}{cc}
M_{GBG} & 0 \\
M_{GBE}^T  & M_{EBE} - M_{GBE}^T M_{GBG}^{-1} M_{GBE}
\end{array} \right) \left( \begin{array}{cc}
I & M_{GBG}^{-1} M_{GBE} \\
0  & I
\end{array} \right). \nonumber
\end{align}
By introducing the pressure prediction value $\phi^*$ and using the right-hand side from the original equation (\ref{eq:NewtMat-proj2}), we can outline steps for solving the modified pressure Poisson equation. First, we solve for $\phi^{*}$ and $\Gamma$
\begin{align}
M_{GBG} \,\phi^{*} & = G^T q^{*} + bc_2, \label{eq:second-blockLU-proj1a}\\
M_{GBG} \,\Gamma & = M_{GBE}, \label{eq:second-blockLU-proj1b}
\end{align}
where the matrix $\Gamma$ is of size $np \times nb$, where $np$ is the number of pressure nodes. In the second step, we factorize $\Gamma$, i.e. $\Gamma = M_{GBG}^{-1} M_{GBE}$. We note that in this step we solve $nb + 1$ independent linear systems with constant (time independent) positive definite matrices $M_{GBG}$. The matrix $M_{GBG}$ factors can be precomputed (using, for example, Cholesky factorization) outside the time loop, and used to find solution for $\phi^{*}$ and $\Gamma$ in parallel.  While this approach suggests to solve more pressure Poisson equations in total, the advantage is time independent factors and the ability to solve $nb + 1$ equations in parallel for the same time step, without relying on partitioning sparse matrices or fluid domain. In the third step we solve for IB forcing
\begin{equation}
\left( M_{EBE} - M_{GBE}^T\, \Gamma \right) \tilde{f}^{n+1/2} = - M_{GBE}^T  \phi^{*} + E_n q^{*} + N_B^T u_B^{*} - \Delta u_B^{n+1}, \label{eq:second-blockLU-proj2}
\end{equation}
where $\left( M_{EBE} - M_{GBE}^T\, \Gamma \right)$ is a dense matrix of size $nb \times nb$, which can be solved efficiently on parallel machines~\cite{slug}. The fourth and final step is projection of the pressure variable
\begin{equation}
\phi^{n+1/2} = \phi^{*} - \Gamma\, \tilde{f}^{n+1/2},
\end{equation}
which involves only matrix multiplication. We note that the second block-LU decomposition can be also applied to non-positive-definite formulation (see Appendix. Stability of the implicit coupling for massless particles) and would give similar time independent positive-definite systems as equations (\ref{eq:second-blockLU-proj1a}--\ref{eq:second-blockLU-proj1b}). The decomposition would also be applicable, if the equation system would be complemented with more complex equations.

We emphasize that the current decomposition allows factorization of linear systems outside of time loop -- thus saving computational time -- and that the problem of a general dense matrix scale better on parallel machines compared to general sparse matrix.

\section*{References}

\bibliography{references_UgisL}

\end{document}